\overfullrule=0pt
\input harvmac
\def\bar{\overline}
\def\a{{\alpha}}
\def\m{{\mu}}

\def\md{{\dot \mu}}
\def\ah{{\widehat \a}}
\def\Ah{{\widehat A}}
\def\ww{{\widehat w}}

\def\dh{{\widehat d}}
\def\lh{{\widehat \lambda}}
\def\muh{{\widehat \mu}}

\def\wh{{\widehat w}}

\def\l{{\lambda}}

\def\b{{\beta}}
\def\bh{{\widehat\beta}}
\def\dh{{\widehat\delta}}
\def\g{{\gamma}}
\def\gh{{\widehat\gamma}}

\def\d{{\delta}}

\def\s{{\sigma}}

\def\half{{1\over 2}}
\def\p{{\partial}}
\def\ads{{{$AdS_5\times S^5$}}}

\def\pb{{\overline\partial}}
\def\t{{\theta}}
\def\md{{\dot\mu}}

\def\th{{\widehat\theta}}

\def\tb{{\bar\theta}}

\Title{\vbox{\baselineskip12pt
\hbox{IFT-P.022/2008 }}}
{{\vbox{\centerline{ Simplifying and Extending the
 }
\smallskip
\centerline{$AdS_5\times S^5$ Pure Spinor Formalism}}} }
\bigskip\centerline{Nathan Berkovits\foot{e-mail: nberkovi@ift.unesp.br}}
\bigskip
\centerline{\it Instituto de F\'\i sica Te\'orica, S\~ao Paulo State University 
}
\centerline{\it Rua Pamplona 145, 01405-900, S\~ao Paulo, SP, Brasil}
\bigskip

\vskip .3in

Although the $AdS_5\times S^5$ worldsheet action is not quadratic,
some features of the pure spinor formalism are simpler in an
$AdS_5\times S^5$ background than in a flat background. The BRST
operator acts geometrically, the left and right-moving pure 
spinor ghosts can be treated as complex conjugates, the zero mode
measure factor is trivial, and the $b$ ghost does not require non-minimal
fields. 

Furthermore, a topological version of the $AdS_5\times S^5$
action with the same worldsheet variables
and BRST operator can be constructed by gauge-fixing
a $G/G$ principal chiral model where 
$G=PSU(2,2|4)$. This topological model is argued to describe
the zero radius limit that is dual to free ${\cal N}=4$ super-Yang-Mills
and can also be interpreted as an ``unbroken phase'' of superstring theory.

\vskip .3in

\Date {December 2008}

\newsec{Introduction}

Up to now, the only superstring formalism suitable for covariantly
quantizing the $AdS_5\times S^5$ background is the pure spinor
formalism
\ref\pureone{N. Berkovits,
{\it Super-Poincar\'e covariant quantization of the superstring},
JHEP 0004 (2000) 018, hep-th/0001035.}.
Because of the
Ramond-Ramond flux, the Ramond-Neveu-Schwarz formalism cannot describe
this background. Although the covariant Green-Schwarz formalism can
classically describe the $AdS_5\times S^5$ background, this formalism
has only been quantized in light-cone gauge by expanding around
classical solutions which break the target-space $PSU(2,2|4)$
invariance. It should be noted that
for computing the physical spectrum, the light-cone Green-Schwarz
formalism is probably the most convenient since it includes only
physical degrees of freedom and does not require ghosts.
However, for computing scattering amplitudes or for describing the spectrum
in a $PSU(2,2|4)$-invariant manner, the pure spinor formalism is
expected to be more convenient since it manifestly preserves all
symmetries.

In a flat target-space background, the worldsheet action in the
pure spinor formalism is quadratic and it is easy to compute 
scattering amplitudes using the free-field OPE's of the
worldsheet fields. However, in an $AdS_5\times S^5$ background,
the worldsheet action is 
\ref\adsqt{N. Berkovits,
{\it Quantum consistency of the superstring in $AdS_5\times S^5$
background}, JHEP 0503 (2005) 041, hep-th/0411170.}
\eqn\adsactp{S = \int d^2 z [
\half \eta_{ab} J^a \bar J^b - \eta_{\a\bh}
({3\over 4}J^\bh \bar J^\a  + {1\over 4}
\bar J^\bh  J^\a) -
w_\a  \bar\nabla \lambda^\a  +
\widehat w_\ah \nabla \lh^\ah  -
{1\over 4} \eta_{[ab][cd]}(w\g^{ab}\l)(\ww \g^{cd}\lh) ]}
where
$J^A= (g^{-1}\p g)^A$ and $\bar J^A = (g^{-1}\pb g)^A$
are the Metsaev-Tseytlin
left-invariant currents
\ref \metsaev{R.R. Metsaev
and A.A. Tseytlin, {\it Type IIB Superstring Action in 
$AdS_5\times S^5$
Background}, Nucl. Phys. B533 (1998) 109, hep-th/9805028.},
$A=(a,\a,\ah,[ab])$ are the
$PSU(2,2|4)$ Lie-algebra indices,
$g$ takes values in the ${{PSU(2,2|4)}\over{SO(4,1)\times SO(5)}}$ coset,
$(\l^\a,w_\a)$
and $(\lh^\ah, \ww_\ah)$ are the left and right-moving pure spinor
variables,
and 
($\eta_{ab}$,$\eta_{\a\bh}$,$\eta_{[ab][cd]}$)
are the nonvanishing components of the $PSU(2,2|4)$
metric.
The global
$PSU(2,2|4)$ isometries act on $g$ by left multiplication 
as $\d g= \Sigma g$, and these global isometries commute
with the BRST transformations
which act by right multiplication as
\eqn\Qaction{Q g = g ~(\l^\a T_\a + \lh^\ah T_\ah)}
where $T_\a$ and $T_\ah$ are the fermionic generators of $PSU(2,2|4)$.
Since the $J^A$ currents are 
not holomorphic, it is difficult to compute OPE's and
scattering amplitudes in an \ads\ background.

Nevertheless, it will be shown in the first half of this paper
that there are several features of the pure spinor formalism in
an \ads\ background which are simpler than in a flat background.
Unlike the worldsheet Lagrangian in a flat background which transforms
by a total derivative under $d=10$ supersymmetry transformations, 
the worldsheet Lagrangian of \adsactp\ is manifestly
$PSU(2,2|4)$ invariant. As a consequence, the vertex operator
for the zero-momentum dilaton in an \ads\ background is 
manifestly $PSU(2,2|4)$ invariant and can be expressed as
the ghost-number $(1,1)$ operator
\eqn\vdil{V^{AdS} = \eta_{\a\ah} \l^\a \lh^\ah}
where $\eta_{\a\ah} \equiv (\g^{01234})_{\a\ah}$.
On the other hand, the zero-momentum dilaton
vertex operator in a flat background is 
\eqn\vflat{V^{flat} = 
(\l\g^m\t)(\lh\g_m\th),}
which transforms under spacetime supersymmetry
into a BRST-trivial operator.

Because $(\eta_{\a\ah}\l^\a\lh^\ah)$ is in the BRST cohomology in
an \ads\ background, it is consistent to impose the
constraint that $(\eta\l\lh)$ is non-vanishing
and to extend the Hilbert space
to include states which depend on inverse powers of $(\eta\l\lh)$.
Note that in a flat background, $(\eta\l\lh)$
is not in the cohomology and can be written as $(\eta\l\lh)=Q(
\eta_{\a\ah}\t^\a\lh^\ah)$. So in a flat background, such an extension
of the Hilbert space would trivialize the cohomology because of the state 
$W= (\eta\l\lh)^{-1} \eta_{\b\bh}\t^\b\lh^\bh$ satisfying $QW=1$, which
would imply that any BRST-closed state $V$ could be written as $V=Q(WV)$.

After extending the Hilbert space in this manner and interpreting
$\l^\a$ and $\eta_{\a\ah}\lh^\ah$ as complex conjugates, it is straightforward
to define functional integration over the pure spinor variables.
Unlike in a flat background where one needs to introduce additional
``non-minimal'' variables to functionally integrate over pure spinors
\ref\ntwo{N. Berkovits,
{\it  Pure spinor formalism as an N=2 topological string},
JHEP 0510 (2005) 089, hep-th/0509120.} 
\ref\nekrb{N. Berkovits
and N. Nekrasov, {\it
Multiloop superstring amplitudes from non-minimal pure spinor formalism},
JHEP 0612 (2006) 029, hep-th/0609012.},
there is no need to introduce non-minimal variables in an \ads\
background. In some sense, the non-holomorphic structure of the \ads\
sigma model automatically regularizes the $0/0$ divergences which
were regularized in a flat background by the non-minimal variables.

Since there are no non-minimal variables, the zero mode measure
factor and the composite $b$ ghost are simpler in an \ads\ background
than in a flat background. In a flat background, the tree-level zero
mode measure factor is
\eqn\zmode{\langle f(x,\t,\l,\widehat\t,\lh)\rangle = 
\int d^{10}x \int (d^5\t)_{\a_1 ... \a_5}
(d^5\th)_{\ah_1 ... \ah_5} }
$$
 (\g^m {\p\over{\p\l}})^{\a_1}
 (\g^n {\p\over{\p\l}})^{\a_2}
 (\g^p {\p\over{\p\l}})^{\a_3}
 (\g_{mnp})^{\a_4\a_5} 
 (\g^q {\p\over{\p\lh}})^{\ah_1}
 (\g^r {\p\over{\p\lh}})^{\ah_2}
 (\g^s {\p\over{\p\lh}})^{\ah_3}
 (\g_{qrs})^{\ah_4\ah_5} $$
$$
f(x,\t,\l,\widehat\t,\lh)|_{\t=\th=0} $$
and the $b$ ghost satisfying $\{Q,b\}=T$ depends in a complicated
manner on the non-minimal variables. In an \ads\ background, the
tree-level zero mode measure factor is simply
\eqn\zmodet{
\langle f(x,\t,\l,\th,\lh)\rangle = 
\int d^{10}x\int d^{16}\t d^{16}\th 
~sdet(E_M^A) \int d\l d\lh ~ f(x,\t,\l,\th,\lh)}
where $E_M^A$ is the target-space supervierbein and $\int d\l d\lh$
is a compact integration over the projective pure spinors.
And the composite $b$ ghost is
\eqn\bg{b = (\eta\l\lh)^{-1} ~\lh^\ah
[\half(\g_a J)_\ah J^a 
+ {1\over 4}\eta_{\a\ah} N^{ab} (\g_{ab}J)^\a + {1\over 4}
\eta_{\a\ah} J_{gh} J^\a]}
where $(J^\a, J^a, J^\ah)$ are the left-invariant currents
constructed from $g$,
and $N^{ab}$ and $J_{gh}$ are the
Lorentz and ghost-currents for $\l^\a$. 

It is instructive to consider the pure spinor formalism
for the Ramond-Ramond plane-wave background 
\ref\rrp
{N. Berkovits, {\it Conformal field theory for the superstring
in a Ramond-Ramond plane wave background}, JHEP 0204 (2002) 037,
hep-th/0203248.} 
where a partial
simplification also occurs. In this background, the operator of \vdil\
is replaced with $(\l\g_{+1234}\lh)$ which only involves the
$(\g_+\l)$ and $(\g_+\lh)$ components of the pure spinors.
So one still needs to introduce non-minimal variables for the
$(\g_-\l)$ and $(\g_-\lh)$ components in order
to perform functional integration. This implies that the
tree-level measure factor in the plane-wave background involves integration
over 18 $\t$'s, as opposed to the 10 $\t$'s in a flat background
or the 32 $\t$'s in an \ads\ background.

In principle, these results could be used to compute \ads\ scattering
amplitudes without the regularization complications that plague
amplitude computations in a flat background \ntwo\nekrb. 
Unfortunately, the difficulties with evaluating OPE's and with
constructing explicit vertex operators
in an \ads\ background will probably make it hard to compute non-trivial
scattering amplitudes at finite $AdS$ radius. Nevertheless, it
might eventually be possible to compute amplitudes 
at infinitesimally small $AdS$ radius and test the
Maldacena conjecture in the perturbative super-Yang-Mills regime.

In order to compute superstring amplitudes in this perturbative
super-Yang-Mills regime, the first step would
be construct a closed string theory that describes
the zero radius limit that is dual to free ${\cal N}=4$
super-Yang-Mills theory \ref\gopak
{R. Gopakumar, {\it From free fields to AdS}, Phys. Rev. D70 (2004) 025009,
hep-th/0308184.}. Since super-Yang-Mills is a field theory,
it is natural to try to describe this zero radius limit using
a topological string theory \ref\witten
{E. Witten, {\it Perturbative gauge theory as a string theory in twistor space},
Comm. Math. Phys. 252 (2004) 189, hep-th/0312171.}. 
One recent topological string proposal 
\ref\topone{N. Berkovits,
{\it A new limit of the $AdS_5\times S^5$ sigma model}, JHEP 0708 (2007) 011, 
hep-th/0703282.}
\ref\limtwo{N. Berkovits and C. Vafa,
{\it Towards a worldsheet derivation of the Maldacena conjecture},
JHEP 0803 (2008) 031, arXiv:0711.1799[hep-th].}
was 
constructed from the fermionic coset ${{PSU(2,2|4)}\over{SO(4,2)\times
SO(6)}}$ which was related by a field redefinition to the pure spinor
formalism.
This topological string theory was later obtained in 
\ref\topthree{N. Berkovits, 
{\it Perturbative super-Yang-Mills from the topological
$AdS_5\times S^5$ sigma model}, JHEP 0809 (2008) 088,
arXiv:0806.1960[hep-th].} 
by gauge-fixing the ${ G}/{ G}$ principal chiral model
with ${ G}=PSU(2,2|4)$, and similar ${ G}/{ G}$ topological
models for the zero radius limit
have been proposed by A. Polyakov\ref\poly{A. Polyakov, {\it Old and
new aspects of the strings/gauge correspondence}, Strings 2002
proceedings, http://www.damtp.cam.ac.uk/strings02/avt/polyakov/.}
and H. Verlinde\ref\verlin{H. Verlinde, private communication.}.

In the second half of this paper, it will be shown that there is
an alternative gauge-fixing of the ${ G}/{ G}$ principal
chiral model which produces a topological string theory based on
the Metsaev-Tseytlin coset ${{PSU(2,2|4)}\over{SO(4,1)\times SO(5)}}$
instead of the fermionic coset 
${{PSU(2,2|4)}\over{SO(4,2)\times SO(6)}}$.
This alternative gauge-fixing is related to an \ads\ generalization
of the ``extended pure spinor'' formalism proposed by Aisaka and Kazama 
\ref\aisaka{Y. Aisaka and Y. Kazama, {\it A new first class algebra,
homological perturbation and extension of pure spinor formalism
for superstring}, JHEP 0302 (2003) 017, hep-th/0212316.} and,
unlike the BRST transformation for the gauge-fixing to the fermionic
coset, the BRST transformation using this alternative gauge-fixing
is the same as in \Qaction. 

The worldsheet action of this topological string theory is BRST-trivial
and is
\eqn\topact{S_{top} = \int d^2 z [{{(\l\g_a\g_b\lh)}\over{2(\eta\l\lh)}}
J^a \bar J^b + \eta_{\a\ah} \bar J^\a J^\ah  - w_\a \bar\nabla
\l^\a + \widehat w_\ah \nabla \lh^\ah -{1\over 4} \eta_{[ab][cd]}
(w\g^{ab}\l)(\ww\g^{cd}\lh)],}
where $J^A = (g^{-1}\p g)^A$ are the same left-invariant currents constructed
from a ${{PSU(2,2|4)}\over{SO(4,1)\times SO(5)}}$ coset as before.
Note 
that \topact\ differs from the original \ads\ action of \adsactp\
through the $(\l^\a,\lh^\ah)$ dependence of the first term and
the absence of an $\eta_{\a\ah} J^\a \bar J^\ah$ term.

To show that this topological string theory is the dual to free
${\cal N}=4$ super-Yang-Mills, the first step is
to show that the BRST cohomology correctly reproduces the single-trace
gauge-invariant super-Yang-Mills operators at zero `t Hooft
coupling. Since the topological BRST
transformations are the same as in the original \ads\ model, it is
trivial to show that vertex operators for
half-BPS states in the original \ads\ sigma model are also in the BRST
cohomology of the topological sigma model.
Vertex operators for non-BPS states can 
be constructed by 
acting on half-BPS vertex operators with
the $\s$-dependent transformation
\eqn\translo{\d g(\s) = \Sigma(\s) g(\s)} 
where $\Sigma(\s)$
is an arbitrary local $PSU(2,2|4)$ transformation whose $\s$-independent
modes are the global isometries. These transformations commute
with the BRST transformations of \Qaction, and when acting on
operators of large $R$-charge, the $\s$-dependent modes of $\Sigma$ 
act like
the massive string modes in a plane-wave background by inserting
``impurities'' in the long operator \ref\bmn{D. Berenstein, J. Maldacena
and H. Nastase, {\it Strings in flat space and pp waves 
from N=4 super-Yang-Mills}, JHEP 0204 (2002) 013, hep-th/0202021.}. 
Although the $\s$-dependent
transformations of \translo\
do not leave invariant the topological action of \topact, 
they only change \topact\ by a BRST-trivial term.

The next step to showing that this
topological string theory describes free
${\cal N}=4$ super-Yang-Mills is to show that the topological
string amplitudes correctly reproduce super-Yang-Mills 
amplitudes in the limit of small `t Hooft coupling. For string tree amplitudes
involving three half-BPS states, these amplitudes are guaranteed to
agree since the zero mode measure factor in the topological theory
is the same as in \zmodet\ and since these three-point BPS amplitudes
do not depend on the $AdS$ radius. 

To show the equivalence of other
types of amplitudes, a handwaving argument based on
open-closed topological duality will be presented which
will hopefully be made more rigorous in the future.
The argument follows the proposals of \ref\gaiotto{D. Gaiotto
and L. Rastelli, {\it A paradigm of open/closed duality:
Liouville D-branes and the Kontsevitch model}, JHEP 0507 (2005) 053,
hep-th/0312196.} and  
\ref\gopa{R. Gopakumar and C. Vafa, {\it On the gauge theory/gravity
correspondence}, Adv. Theor. Math. Phys. 3 (1999) 1415, hep-th/9811131.}
\ref\oog{H. Ooguri and C. Vafa, {\it Worldsheet derivation of a large $N$
duality}, Nucl. Phys. B641 (2002) 3, hep-th/0205297.} and
uses that the open string field theory obtained
by putting $D_3$ branes at the $AdS_5$ boundary of
the topological string reproduces 
${\cal N}=4$ super-Yang-Mills field theory.
Furthermore, it will be argued that perturbing the closed topological
action of \topact\ by the vertex operator of \adsactp\ as
\eqn\pertar{S_{top}\to S_{top} + r^2 S}
is equivalent to shifting the `t Hooft coupling constant
of the Yang-Mills theory.

In addition to providing a string dual to free super-Yang-Mills,
this topological string also describes an
unbroken phase of closed superstring theory in which all background
fields (including the metric) are treated on the same footing.
Up to BRST-trivial terms, the topological action of \topact\ is independent
of any specific choice for the spacetime
metric, which was one of the original motivations of Witten
for studying topological string theory \ref\wittop{E. Witten,
{\it Topological quantum field theory}, Comm. Math. Phys. 117 (1988)
353.}\ref\wittopt{E. Witten, {\it Topological sigma models},
Comm. Math. Phys. 118 (1988) 411.}\ref\wittopq{E. Witten, 
{\it Spacetime and topological orbifolds}, Phys. Rev. Let. 61 (1988)
670.}.
To recover non-topological backgrounds, one gives expectation
values to the physical moduli of the topological string. For
example, the \ads\ background at nonzero radius is obtained
by perturbing with the physical vertex operator of \adsactp\
for the radius modulus, and
other string theory backgrounds which
are asymptotically \ads\ can be obtained
by perturbing with vertex operators corresponding to other physical moduli.

As in previous topological proposals of Witten for an unbroken phase
of string theory, the
target spacetime in the topological sigma model
requires a complex structure \wittopt\wittopq. But unlike in
previous proposals, the complex structure of spacetime 
is now dynamical and is determined by the pure spinor ghost variables
$\l^\a$ and $\lh^\ah$ which choose a $U(5)$ subgroup of (Wick-rotated)
$SO(10)$.\foot{Similar observations on pure spinors and topological
strings have been made by N. Nekrasov\ref\nekp{N. Nekrasov, private
communication.}.} This can be seen from the kinetic term for the ten $x$'s
in the first term of \topact\ which, to quadratic order, is
$\int d^2 z (2\eta\l\lh)^{-1}(\l\g_a\g_b\lh) \p x^a \pb x^b.$

In section 2 of this paper, the pure spinor version of
the \ads\ sigma model will be reviewed. In section 3, it will
be shown 
that non-minimal variables are unnecessary in this model,
that the zero mode measure factor and
$b$ ghost are much simpler than in a flat background, and
that a partial simplification also occurs in the Ramond-Ramond
plane-wave background.
In section 4,
a BRST-trivial version of the \ads\ sigma model will be constructed
by gauge-fixing a ${ G}/{ G}$ principal chiral model,
and this topological model will be argued to describe
the dual of free super-Yang-Mills. In section 5, conclusions
and open problems
will be discussed.

\newsec{Review of \ads\ Sigma Model}

The pure spinor version of the worldsheet action for the \ads\
superstring can be derived either by constructing the pure spinor
action in a general curved background 
\ref\howe{N. Berkovits and P. Howe, {\it
Ten-dimensional supergravity constraints from the 
pure spinor formalism for the superstring}, Nucl. Phys. B635
(2002) 75, hep-th/0112160.}
and setting the background
superfields to their \ads\ values, or by adding terms to the 
Green-Schwarz \ads\ action which replace $\kappa$ symmetry
with BRST invariance \ref\tonin{I. Oda and M. Tonin,
{\it On the Berkovits covariant quantization of GS superstring},
Phys. Lett. B520 (2001) 398, hep-th/0109051.}.
The second approach is more direct and
will be reviewed here. The structure of supergravity vertex operators
will then be discussed.

\subsec{Green-Schwarz worldsheet action}

In a general Type II supergravity background, the Green-Schwarz
action is 
\eqn\gsac{\int d^2 z \half (G_{MN}(Z)+B_{MN}(Z)) \p Z^M \pb Z^N = 
\int d^2 z \half (\eta_{ab} E_M^a(Z) E_N^b(Z) + B_{MN}(Z)) \p Z^M \pb Z^N}
where $Z^M = (x^m,\t^\mu,\th^\muh)$, $E_M^A(Z)$ is
the super-vierbein, $A=(a,\a,\ah)$ are tangent-superspace variables
for $a=0$ to 9 and $\a,\ah=1$ to 16,
and $M= (m,\mu,\muh)$ are coordinate variables for $m=0$ to 9 and
$\mu,\muh=1$ to 16, and $(\a,\mu)$ and $(\ah,\muh)$ label spinors
of the opposite/same chirality for the Type IIA/B superstring.

In an $AdS_5\times S^5$ background, the supervierbein $E_M^A$
can be explicitly constructed 
from the Metsaev-Tseytlin
left-invariant
currents
$J^{\tilde A} = (g^{-1} \p g)^{\tilde A}$ where
$g$
takes values in the coset $PSU(2,2|4)/(SO(4,1)\times SO(5))$,
$\tilde A= ([ab],a,\a,\ah)$ ranges over the
30 bosonic and 32 fermionic elements in the Lie algebra of $PSU(2,2|4)$,
$[ab]$ labels the $SO(4,1)\times SO(5)$ ``Lorentz''
generators, $a=0$ to 9 labels
the ``translation'' generators, and $\a,\ah=1$ to 16 label the
fermionic ``supersymmetry'' generators.
Note that $\tilde A$ includes both the superspace indices $A$ as
well as the $SO(4,1)\times SO(5)$ indices $[ab]$. 
The $PSU(2,2|4)$ structure constants $f_{\tilde A \tilde B}^{\tilde C}$
include $f_{\a\b}^a = \g_{\a\b}^a$
and $f_{\ah\bh}^a = \g_{\ah\bh}^a$ where $\g^a_{\a\b}$ and $(\g^a)^{\a\b}$
are the $16\times 16$ off-diagonal elements in the Weyl representation
of the $32\times 32$ ten-dimensional $\Gamma$-matrices, and
$\g^a_{\ah\bh}$ and $(\g^a)^{\ah\bh}$ are related to these matrices by
\eqn\relatig{\g^a_{\ah\bh} \equiv \eta_{\a\ah}\eta_{\b\bh} (\g^a)^{\a\b},
\quad(\g^a)^{\ah\bh} \equiv \eta^{\a\ah}\eta^{\b\bh} \g^a_{\a\b},\quad
\eta_{\a\bh} \equiv (\g^{01234})_{\a\bh}, 
\quad
\eta^{\a\bh} \equiv (\g^{01234})^{\a\bh} .}

Parameterizing the \ads\ coset as
\eqn\gparam{g(Z) = \exp (x^m P_m + \t^\mu Q_\mu + \th^\muh \widehat Q_\muh)}
where $[P_m, Q_\mu, \widehat Q_\muh]$ are the \ads\ translation and
supersymmetry generators, one obtains
\eqn\jcur{J^A = E^A_M(Z) \p Z^M, \quad J^{[ab]} = \omega_M^{[ab]}(Z)\p Z^M}
where $\omega_M^{[ab]}$ is the \ads\ spin connection.
Furthermore, in an \ads\ background, it was shown in \ref\bersh
{N. Berkovits, M. Bershadsky, T. Hauer, S. Zhukov and B. Zwiebach,
{\it Superstring theory on $AdS_2\times S^2$ as a coset supermanifold},
Nucl. Phys. B567 (2000) 61, hep-th/9907200.}
that the only nonzero components
of $B_{AB}= E_A^M E_B^N B_{MN}$ are 
\eqn\nonzc{B_{\a\bh} = B_{\bh\a} =\half (\g^{01234})_{\a\bh} \equiv 
\half\eta_{\a\bh}.}
So the Green-Schwarz action in an \ads\ background is \metsaev\bersh
\eqn\sgs{S_{GS} = \int d^2 z (\half \eta_{ab} J^a \bar J^b + 
{1\over 4} \eta_{\a\bh} (J^\a \bar J^\bh - \bar J^\a J^\bh)) .}
Note that unlike the Green-Schwarz Lagrangian in a flat background
in which the term $B_{MN}\p Z^M \pb Z^N$ transforms by a total
derivative under spacetime supersymmetry, the Green-Schwarz
Lagrangian in an \ads\ background is manifestly $PSU(2,2|4)$
invariant since it can be expressed in terms of the supersymmetric
invariants $J^A$.

\subsec{Pure spinor worldsheet action}

To generalize the Green-Schwarz action to the pure spinor
formalism, one needs to add canonical momenta $(d_\a,\widehat d_\ah)$
for the $(\t^\mu,\th^\muh)$ variables as well as left
and right-moving pure spinor ghosts,
$(\l^\a, w_\a)$ and $(\lh^\ah,\wh_\ah)$, 
which satisfy the pure spinor constraints
$\l\g^a\l=\lh\g^a\lh=0$. Because of the pure spinor constraints,
$w_\a$ and $\widehat w_\ah$ can only appear in combinations which are
invariant under the gauge transformations
\eqn\wwgauge{\d w_\a= \xi^a (\g_a\l)_\a, \quad
\d \widehat w_\ah= \widehat\xi^a (\g_a\lh)_\ah,}
which implies
that they only appear through the Lorentz currents and ghost currents
\eqn\ghcurr{N_{ab}=\half w\g_{ab}\l,\quad J_{gh} = w_\a\l^\a,\quad\quad
\widehat N_{ab}=\half \widehat w\g_{ab}\lh,\quad \widehat J_{gh}
=\widehat w_\ah\lh^\ah.}

In an \ads\ background, these additional worldsheet fields couple as
\eqn\adsp{S = S_{GS} + \int d^2 z [- d_\a \bar J^\a + \widehat d_\ah
J^\ah + d_\a \widehat d_\bh F^{\a\bh} - 
w_\a (\bar\nabla \l)^\a + 
\widehat w_\ah (\nabla \lh)^\ah + R_{abcd}N^{ab}\widehat N^{cd}]}
where $F^{\a\bh} = (\g_{01234})^{\a\bh}\equiv
\eta^{\a\bh}$ is the bispinor Ramond-Ramond field-strength, 
$R_{abcd}=\mp \eta_{a[c}\eta_{d]b}\equiv -
\eta_{[ab][cd]}$ is the \ads\ curvature
(the $-$ sign is if $a,b,c,d$ are on $AdS_5$ and the $+$ sign is
if they are on $S^5$), and 
\eqn\deldef{(\bar\nabla\l)^\a = \pb\l^\a + \half\bar J^{[ab]}(\g_{ab}\l)^\a,
\quad
(\nabla\lh)^\ah = \p\lh^\ah + \half J^{[ab]}(\g_{ab}\lh)^\ah.}

Because of the nonvanishing Ramond-Ramond flux, $d_\a$ and $\widehat d_\ah$
are auxiliary fields which can be integrated out to give the action
\eqn\class{S = \int d^2 z [\half \eta_{ab} J^a \bar J^b - \eta_{\a\bh}
({3\over 4}J^\bh \bar J^\a  + {1\over 4}
\bar J^\bh  J^\a) -
w_\a  \bar\nabla \lambda^\a  +
\widehat w_\ah \nabla \lh^\ah  -
 \eta_{[ab][cd]}N^{ab} \widehat N^{cd} ]}
\eqn\classtwo{=
\int d^2 z[ \half (\eta_{ab} J^a \bar J^b + \eta_{\a\bh} J^\a \bar J^\bh
+\eta_{\a\bh} \bar J^\a J^\bh) -{1\over 4} \eta_{\a\bh}
(J^\a\bar J^\bh - \bar J^\a J^\bh)}
$$
+ (- w_\a\bar\nabla\l^\a + \widehat w_\ah\nabla\lh^\ah -
\eta_{[ab][cd]}
N^{ab} \widehat N^{cd} )]. $$
The action of \class\ is manifestly invariant under global $PSU(2,2|4)$
transformations which transform $g(x,\t,\th)$ by left multiplication
as $\d g = (\Sigma^{\tilde A} T_{\tilde A}) g$ where
$T_{\tilde A}$ are the $PSU(2,2|4)$ Lie-algebra generators
and is also manifestly invariant
under local $SO(4,1)\times SO(5)$ gauge transformations which transform
$g(x,\t,\th)$ by right multiplication as $\d_\Lambda
g= g(\Lambda^{[ab]}T_{[ab]})$ and transform
the pure spinors as $SO(4,1)\times SO(5)$ target-space spinors.

The BRST operator in the pure spinor formalism is defined as
\eqn\brstp{Q = \int dz ~\l^\a d_\a + \int d\bar z ~\lh^\ah \widehat d_\ah =
 \int dz ~\eta_{\a\ah} \l^\a J^\ah + \int d\bar z ~\eta_{\a\ah}
\lh^\ah \bar J^\a,}
where the auxiliary equations of motion for $d_\a$ and $\widehat d_\ah$
have been used.
Under 
BRST transformations generated by $Q$,
$g(x,\t,\th)$ transforms by right-multiplication as
\eqn\brstt{ Q (g) = g (\l^\a T_\a +\lh^\ah T_\ah)}
which implies that
\eqn\jtrans{QJ^\a = \nabla \l^\a - \eta^{\a\ah}(\g_a\lh)_\ah J^a,\quad
QJ^\ah = 
\nabla \lh^\ah + \eta^{\a\ah}(\g_a\l)_\a J^a,}
$$QJ^a = (\g_a\l)_\a J^\a + (\g_a\lh)_\ah J^\ah, \quad
Q J^{[ab]} = \half \eta^{[ab][cd]} \eta_{\a\ah}
(J^\ah (\g_{cd}\l)^\a - J^\a (\g_{cd}\lh)^\ah).$$
And \brstp\ implies that the pure spinors transform as
\eqn\brstw{Q(w_\a) =  \eta_{\a\ah} J^\ah ,\quad
Q(\widehat w_\ah) = \eta_{\a\ah}\bar J^\a, \quad
Q(\l^\a)= Q(\lh^\ah) =0.}

To verify that \class\
is BRST invariant,
note that the first term in the Lagrangian of 
\classtwo\ transforms under \brstp\ to
$$\half\eta_{\a\ah} (J^\ah \bar\nabla \lambda^\a +\bar J^\ah \nabla \lambda^\a
- J^\a \bar\nabla\lh^\ah - \bar J^\a \nabla\lh^\ah).$$
Using the
Maurer-Cartan equations
\eqn\mceq{\nabla \bar J^\ah - \bar\nabla J^\ah = \g_a^{\ah\bh}\eta_{\b\bh}
(J^\b \bar J^a - \bar J^\b J^a), 
\quad
\nabla \bar J^\a - \bar\nabla J^\a = -\g_a^{\a\b}\eta_{\b\bh}
(J^\bh \bar J^a - \bar J^\bh J^a),}
the second term in \classtwo\ transforms under \brstp\ to
\eqn\surface{
\half\eta_{\a\ah}(J^\ah \bar\nabla\lambda^\a -\bar J^\ah \nabla\lambda^\a
+ J^\a\bar\nabla\lh^\ah - \bar J^\a \nabla\lh^\ah)}
$$ +
{1\over 4}\eta_{\a\ah}\p(\bar J^\ah \l^\a +\bar J^\a \lh^\ah)
-{1\over 4}\eta_{\a\ah}\bar\p(J^\ah \l^\a + J^\a\lh^\ah) .$$
And the last term
in \classtwo\ transforms under \brstp\ to
$$ - \eta_{\a\ah} (J^\ah \bar\nabla\lambda^\a -
\bar J^\a \nabla\lh^\ah).$$
So ignoring the total derivatives in the second line
of \surface, \class\ is BRST-invariant.

\subsec{Nilpotent BRST transformations}

Although it is consistent to use the BRST transformations of
\brstt\ and \brstw\
which are nilpotent up to equations of motion, it will be convenient
to include auxiliary antifields in the action so that the BRST
transformations become nilpotent without using equations of motion.
As discussed in \limtwo\ 
and shown independently by G. Boussard\ref\bouss
{G. Boussard, private communication.}, this is easily
done by adding the antifields $w_\a^*$ and $\ww_\ah^*$ to the
\ads\ action of \class\ as 
\eqn\antiact{S \to S + \int d^2 z \eta^{\a\ah} w_\a^*\ww_\ah^*}
where $w_\a^*$ and $\ww_\ah^*$ are auxiliary fermionic spinors which are
constrained to satisfy
\eqn\constrw{\eta_{\a\ah} (w^*\g^a)^\a\lh^\ah = 0, \quad
\eta_{\a\ah} (\ww^*\g^a)^\ah\l^\a = 0,}
and therefore each contain 11 independent fermionic components.

Under the BRST transformations of \brstt\ and \brstw, one finds that
\eqn\qsq{Q^2 g = - g (h^{[ab]}T_{[ab]}),}
$$Q^2 w_\a = \half (\g_{ab}w)_\a h^{[ab]} + (\l\g_a)_\a \xi^a + \eta_{\a\ah}
{{\p L}\over{\p\ww_\ah}},$$
$$Q^2 \ww_\ah = \half (\g_{ab}\ww)_\ah h^{[ab]} + (\lh\g_a)_\ah 
\widehat\xi^a - \eta_{\a\ah}
{{\p L}\over{\p w_\a}},$$
where 
\eqn\whereqsq{ h^{[ab]}=\half \eta_{\a\ah} \l^\a (\g^{ab}\lh)^\ah, \quad
\xi^a = J^a - \eta^{\a\ah} w_\a (\g^a\lh)_\ah,\quad
\widehat\xi^a = -\bar J^a + \eta^{\a\ah} \ww_\ah (\g^a\l)_\a,}
$$
{{\p L}\over{\p\ww_\ah}} = \nabla \lh^\ah -\half \eta_{[ab][cd]}
N^{ab}(\g^{cd}\lh)^\ah,\quad
{{\p L}\over{\p w_\a}} = -\bar\nabla \l^\a -\half \eta_{[ab][cd]}
(\g^{ab}\l)^\a\widehat N^{cd}.$$

When acting on terms which are gauge-invariant with respect to the
local $SO(4,1)\times SO(5)$ transformations and the $(w,\ww)$ gauge
transformations of \wwgauge, the terms in \qsq\ which are proportional
to $(h^{[ab]}, \xi^a, \widehat \xi^a)$
can be ignored. To remove the terms in \qsq\ which are proportional
to the equations of motion 
${{\p L}\over{\p w_\a}}$ and ${{\p L}\over{\p\ww_\ah}}$, one
should modify the BRST transformations of $w_\a$ and $\ww_\ah$ to
\eqn\modqw{Q w_\a = \eta_{\a\ah} J^\ah + w_\a^*,\quad
Q \ww_\ah = \eta_{\a\ah} \bar J^\a + \ww_\ah^*,}
and define the BRST transformation of the antifields $w_\a^*$ and
$\ww_\ah^*$ as 
$$Q w_\a^* = -\eta_{\a\ah}
{{\p L}\over{\p\ww_\ah}}, \quad
Q \ww_\ah^* = \eta_{\a\ah} {{\p L}\over{\p w_\a}}.$$
With the addition of \antiact\ to the action, one can easily
check that these BRST transformation leave the action invariant
and are nilpotent without using equations of motion.

\subsec{Supergravity vertex operators}

In a general curved supergravity background,
physical closed string vertex operators in the pure spinor formalism
are defined as states of ghost-number $(1,1)$ which are
in the BRST cohomology. For massless supergravity states,
these vertex operators only depend on the zero modes of the
worldsheet fields $Z^M =(x^m, \t^\mu,\th^{\widehat\mu})$ as
\eqn\vert{V = \l^\a \lh^\ah A_{\a\ah}(Z^M).}

Under the BRST transformation generated by
$Q=\int dz \l^\a d_\a + \int d\bar z \lh^\ah \widehat d_\ah$,
\eqn\transfz{Q Z^M = \l^\a E_\a^M(Z) + \lh^\ah E_\ah^M(Z)}
where $E_A^M$ is the inverse supervierbein. So
\eqn\transfv{Q V = \l^\a\lh^\ah (\l^\b E_\b^M\ + \lh^\bh E_\bh^M)
\p_M A_{\a\ah}
= (\l^\b \nabla_\b + \lh^\bh \nabla_\bh) (\l^\a \lh^\ah A_{\a\ah})}
where $\nabla_A = E_A^M (\p_M + \omega_M^{[ab]}M_{[ab]})$ is
the covariant derivative and $M^{[ab]}$ are tangent-space Lorentz
generators which act on the spinor indices $\a$ and $\ah$.
Since $\l\g^a\l=\lh\g^a\lh=0$, $QV=0$ implies that $A_{\a\ah}(Z)$
satisfies 
\ref\adsquant{N. Berkovits and O. Chand\'{\i}a,
{\it Superstring Vertex Operators in an $AdS_5\times S^5$ Background},
Nucl. Phys. B596 (2001) 185, hep-th/0009168.}
\eqn\eqA{\g_{abcde}^{\a\g} \nabla_\g A_{\a\bh} = 
\g_{abcde}^{\bh\gh} \nabla_\gh A_{\a\bh} = 0}
for any choice of $[abcde]$. And the gauge transformation
\eqn\gauget{\d V = Q (\l^\a \Omega_\a + \lh^\ah \Omega_\ah)
= (\l^\b\nabla_\b + \lh^\bh\nabla_\bh)
(\l^\a \Omega_\a + \lh^\ah \Omega_\ah)}
implies that $A_{\a\ah}(Z)$ is defined up to the gauge transformation
\eqn\agauge{\d A_{\a\ah} = \nabla_\a \Omega_\ah + \nabla_\ah \Omega_\a}
where $\Omega_\a$ and $\Omega_\ah$ are restricted to satisfy
\eqn\eqOmega{\g_{abcde}^{\a\b} \nabla_\b \Omega_\a = 
\g_{abcde}^{\ah\bh} \nabla_\bh \Omega_\ah= 0}
for any choice of $[abcde]$. 

As shown in \howe,
these equations of motion and gauge invariances
describe an onshell Type II supergravity multiplet. In terms
of the standard supergravity superfields, $A_{\a\ah}(Z)$ is
identified with the spinor-spinor component 
$B_{\a\bh}$ of the two-form $B_{AB} = E_A^M E_B^N B_{MN}$
in the gauge where $(\g_{abcde})^{\a\b} B_{\a\b} = 
(\g_{abcde})^{\ah\bh} B_{\ah\bh} = 0$. The equations of motion
of \eqA\ follow from the superfield constraints 
\eqn\constraints{H_{\a\bh\gh} = H_{\ah\b\g}=0,\quad 
(\g_{abcde})^{\a\b} T_{\a\b}^D = 
(\g_{abcde})^{\ah\bh} T_{\ah\bh}^D = T_{\a\ah}^D =0,}
where 
\eqn\defH{H_{ABC} = E_A^M E_B^N E_C^P \p_{[M} B_{NP)} = 
\nabla_{[A} B_{BC)} + T_{[AB}^D B_{C)D}}
is the three-form field strength and $T_{AB}^D$ is the superspace torsion.
And the gauge transformations of \agauge\ follow from the
gauge transformations $\d B_{MN} = \p_{[M}\Omega_{N)}$ which imply that
$\d B_{AB} = \nabla_{[A} \Omega_{B)} + T_{AB}^C \Omega_C.$

In a flat background, the constraints of \eqA\ can be easily solved
in terms of plane-wave solutions as $A_{\a\bh}(Z)= A_{\a\bh}(k,\t,\th) e^{ikx}$
where $k^2=0$. Furthermore, the holomorphic structure of the sigma model
implies that $A_{\a\ah}(k,\t,\th)$ factorizes into
$A_{\a\ah}(k,\t,\th) = A_\a(k,\t)A_\ah(k,\th)$ where
$A_\a(k,\t)$ is the super-Yang-Mills spinor gauge field satisfying
$(\g_{abcde})^{\a\b} D_\a A_\b =0$ with $D_\a ={\p\over{\p\t^\a}}
+ k_m \g^m_{\a\b}\t^\b$. 

Unfortunately, the non-holomorphic structure
of the \ads\ sigma model does not allow a similar factorization for
$A_{\a\bh}(Z)$ in an \ads\ background. Nevertheless, the fact that $B_{\a\ah}$
has the background value of $\eta_{\a\ah}$ in this background implies that
the $\t=\th=0$ component of $\eta^{\a\ah}A_{\a\ah}(Z)$ is the dilaton.
The other components of $A_{\a\ah}(Z)$ can be determined by acting with
supersymmetry on the dilaton.
 
\newsec{Simplifying the \ads\ Formalism}

In this section, it will be explained that since $(\eta_{\a\ah}\l^\a\lh^\ah)$
is in the BRST cohomology in an \ads\ background, 
there is no need to introduce the non-minimal
variables which are necessary in a flat background to regularize
the functional integral over the pure spinors.
This simplifies the zero mode measure factor and $b$ ghost in an
\ads\ background, and a partial simplification will also occur in
the Ramond-Ramond plane-wave background.

\subsec{BRST cohomology and extended Hilbert space}

To show that $(\eta\l\lh)$ is in the BRST cohomology in an \ads\
background, note that the surface term in \surface\ implies that
\eqn\qtransf{ Q L_{AdS} = \p \bar f - \pb f}
where $L_{AdS}$ is the Lagrangian of \class\ and
\eqn\defff{ f = {1\over 4}\eta_{\a\ah} (\l^\a J^\ah + \lh^\ah J^\a),\quad
\bar f = 
{1\over 4}\eta_{\a\ah} (\l^\a \bar J^\ah + \lh^\ah \bar J^\a).}
Furthermore, since the BRST transformations of \brstt\ and
\modqw\ are nilpotent, \qtransf\
implies that
$Qf = \p V$ and $Q\bar f = \pb V$ for some $V$. One can easily check
for $f$ and $\bar f$ of \defff\
that $V={1\over 4} \eta_{\a\ah} \l^\a \l^\ah$.

Since this procedure relates dimension $(1,1)$ integrated vertex operators
and dimension $(0,0)$ unintegrated vertex operators, $V= (\eta\l\lh)$
is the unintegrated vertex operator associated with 
the \ads\ Lagrangian. And since the $AdS_5$ radius
which multiplies the Lagrangian
is a physical modulus, $(\eta\l\lh)$ must be in the BRST cohomology.
Note that in a flat background, the analogous procedure using the
flat worldsheet Lagrangian produces the physical unintegrated vertex operator
$V= (\l\g^m\t)(\lh\g_m\th)$. 

Since $(\eta\l\lh)$ is in the BRST cohomology, it is consistent
to impose the constraint that $(\eta\l\lh)$ is
non-vanishing. If $\l^\a$ and $\eta_{\a\ah}\lh^\ah$
are interpreted as complex conjugates, this constraint implies
that at least one component of $\l^\a$ must be nonzero.
In the presence
of this constraint, the Hilbert space can be 
extended to include states which depend on inverse powers of
$(\eta \l\lh)$. 

As mentioned in the introduction, such an extension
of the Hilbert space in a flat background
would trivialize the BRST cohomology 
since it would allow the state $W=(\eta\l\lh)^{-1} (\eta_{\b\bh}
\t^\b\lh^\bh)$ which satisfies $QW=1$. But since $(\eta\l\lh)$ is not
BRST-trivial, there is no such $W$ satisfying $QW=1$
that can be constructed in an
\ads\ background.

\subsec{$b$ ghost}

Since $[Q,T]=0$ where 
\eqn\adsstr{T= \half \eta_{ab} J^a J^b + \eta_{\a\ah}J^\a J^\ah - 
w_\a\nabla\l^\a}
is the left-moving stress tensor, one can ask if there exists an operator
$b$ satisfying $\{Q,b\}=T$. Before extending the Hilbert space to
include inverse powers of $(\eta\l\lh)$, such an operator does not exist.
This situation is analogous to the situation in a flat background where,
before introducing non-minimal fields, one cannot construct an operator
$b$ satisfying $\{Q,b\} = T_{flat}$ where
$T_{flat}= \half\p x^m\p x_m - p_\a \p\t^\a - w_\a \p\l^\a$.

However, after extending the Hilbert space to include inverse powers
of $(\eta\l\lh)$, the $b$ operator can be defined as 
\eqn\bnew{b = (\eta\l\lh)^{-1}  \lh^\ah [\half\g_{a\ah\bh} J^a J^\bh 
+ {1\over 4}
(\g_{ab})_\ah{}^\bh \eta_{\b\bh} N^{ab} J^\b +{1\over 4}
 \eta_{\a\ah} J_{gh} J^\a].}
Note that \bnew\ resembles the first term of the
$b$ ghost in a flat background which is 
\ref\rnspure{N. Berkovits, {\it Relating the RNS and pure spinor
formalisms for the superstring}, JHEP 0108 (2001) 026, hep-th/0104247.}
\eqn\bflat{b_{flat} = (\l^\a \bar\l_\a)^{-1} \bar\l_\a [
\half\g_m^{\a\b} \Pi^m d_\b +{1\over 4}
 (\g_{mn})_\b{}^\a N^{mn} \p\t^\b + {1\over 4}J_{gh}\p\t^\a ]
+ ...}
where $\bar\l_\a$ is a non-minimal field and $...$ includes terms with
more complicated dependence on the non-minimal fields.

To show that $\{Q,b\}=T$, use \brstt\ to compute that 
\eqn\bcomput{Qb = 
(\eta\l\lh)^{-1} [\half (\eta\l\lh)\eta_{ab}J^a J^b
+ \half(\l\g_a)_\a J^\a (\lh\g^a)_\ah J^\ah }
$$+{1\over 4} \lh^\ah (\g^{ab})_\ah{}^\bh \eta_{\b\bh} N_{ab} \nabla \l^\b
+{1\over 8}(J^\ah (\g^{ab})_\ah{}^\bh \eta_{\b\bh}\l^\b)
(\lh^\gh (\g^{ab})_\gh{}^\dh \eta_{\d\dh}J^\d)$$
$$+{1\over 4}(\eta_{\a\ah}\l^\a J^\ah)(\eta_{\b\bh}\lh^\bh J^\b)
+{1\over 4} (\eta_{\a\ah}\lh^\ah \nabla\l^\a) J_{gh}]$$
$$=
\half \eta_{ab} J^a J^b + \eta_{\a\ah}J^\a J^\ah - 
w_\a\nabla\l^\a$$
where the identity
\eqn\identt{\d_\a^\d \d_\b^\g =\half (\g^a)_{\a\b} (\g_a)^{\g\d} -{1\over 8}
 (\g^{ab})_\a{}^\g (\g_{ab})_\b{}^\d
-{1\over 4} \d_\a^\g \d_\b^\d }
has been used and terms proportional to $w_\a^*$ have been dropped
since they vanish onshell.
One can similarly define $\bar b$ satisfying $\{Q,\bar b\}= \bar T$
where $\bar T = 
\half\eta_{ab}\bar J^a \bar J^b + \eta_{\a\ah}\bar J^\a \bar J^\ah
 + \widehat w_\ah\bar\nabla\lh^\ah$
and 
\eqn\defbarb{\bar b=
(\eta\l\lh)^{-1}  \l^\a [-\half \g_{a\a\b} \bar J^a \bar J^\b -{1\over 4}
(\g_{ab})_\a{}^\b \eta_{\b\bh} \widehat N^{ab} \bar J^\bh  -{1\over 4}
\eta_{\a\ah} \widehat J_{gh}\bar J^\ah].}

Note that $b$ is not holomorphic but $\pb b$ is BRST-trivial. 
The $g$-loop amplitude prescription in the pure 
spinor formalism is given by
\eqn\gloop{A_g = \int d^{3g-3}\tau \int d^{3g-3}\bar\tau
\langle (\int \mu b)^{3g-3} (\int \bar\mu \bar b)^{3g-3}
\prod_{r=1}^N \int d^2 z_r U_r(z_r)\rangle }
where $U_r$ are the dimension $(1,1)$ integrated vertex operators
and $\mu$ and $\bar\mu$ are the Beltrami differentials associated
with the Teichmuller parameters $\tau$ and $\bar\tau$. One normally
requires $\pb b=0$ so that $(\int \mu b)$ is invariant under
transformations that shift $\mu$ by $\bar\p \nu$ for any $\nu$.
However, assuming that BRST-trivial
terms in the integrand do not contribute, 
it seems to be sufficient to only require that $\pb b$ is
BRST-trivial.

\subsec{Functional integration and measure factor}

In a flat background, functional integration over the 22 
zero modes of $\l^\a$ and $\lh^\ah$ produces a divergent factor
since these bosonic zero modes are non-compact. 
The most convenient method for regularizing this divergence is 
to introduce ``non-minimal'' variables $\bar\l_\a$ and $\bar{\lh}_\ah$,
together with their BRST superpartners $r_\a$ and $\widehat r_\ah$,
and to modify the BRST operator to \ref\pertw{E. Witten,
{\it Two-dimensional models with $(0,2)$ supersymmetry},
hep-th/0504078.}\ntwo\nekrb
\eqn\brstnon{Q_{non-min} = \int dz (\l^\a d_\a + r_\a \bar w^\a) +
\int d\bar z (\lh^\ah \widehat d_\ah + \widehat r_\ah \bar {\widehat w}^\ah) }
where $\bar w^\a$ and $\bar {\widehat w}^\ah$ are the conjugate momenta
for $\bar\l_\a$ and $\bar{\lh}_\ah$ and the non-minimal variables satisfy
the constraints
\eqn\nomincon{\bar\l\g^m\bar\l = \bar\l\g^m r = \bar{\lh}\g^m\bar{\lh}
=\bar{\lh}\g^m \widehat r =0.} 

One then inserts the regulator 
\eqn\regu{{\cal N} =\exp[-\rho~Q(\t^\a \bar\l_\a +\th^\ah\bar{\lh}_\ah)]=
 \exp[-\rho( \l^\a\bar\l_\a + \lh^\ah\bar{\lh}_\ah
-\t^\a r_\a -\th^\ah \widehat r_\ah)]}
into the functional integral where $\rho$ is a positive constant.
Since 
${\cal N} -1$ is BRST-trivial, the amplitude
must be independent of the constant $\rho$ and the location of ${\cal N}$.
Treating $\bar \l_\a$ and $\bar{\lh}_\ah$ as the complex conjugates
of $\l^\a$ and $\lh^\ah$, the insertion of ${\cal N}$ regularizes
the functional integration over the pure spinor ghost zero modes
because of its Gaussian dependence on $\l$. 
As shown in \ntwo,
functional integration using this regularization method
in a flat background implies that 
\eqn\funf{\langle f(x,\t,\l,\th,\lh) \rangle =
\int d^{10}x \int d^{11}\l d^{11}\lh d^{11}\bar\l d^{11}\bar{\lh}
\int d^{16}\t d^{16}\th d^{11}r d^{11}\widehat r~{\cal N}~ f(x,\t,\l,\th,\lh)}
$$=\int d^{10}x \int (d^5\t)_{\a_1 ... \a_5}
(d^5\th)_{\ah_1 ... \ah_5}  $$
$$
 (\g^m {\p\over{\p\l}})^{\a_1}
 (\g^n {\p\over{\p\l}})^{\a_2}
 (\g^p {\p\over{\p\l}})^{\a_3}
 (\g_{mnp})^{\a_4\a_5} 
 (\g^q {\p\over{\p\lh}})^{\ah_1}
 (\g^r {\p\over{\p\lh}})^{\ah_2}
 (\g^s {\p\over{\p\lh}})^{\ah_3}
 (\g_{qrs})^{\ah_4\ah_5} $$
$$ f(x,\t,\l,\widehat\t,\lh)_{\t=\th=0}$$
where $f(x,\t,\l,\th,\lh)$ is assumed to have ghost-number $(3,3)$
and
be independent of
the non-minimal fields.
Note that \nomincon\ 
implies that $r_\a$ and $\widehat r_\ah$ each have 11 independent
components, and integration over these components reduces the 
$\int d^{16}\t d^{16}\th$ integral to $\int d^5\t d^5\th$ because of
the $r_\a$ and $\widehat r_\ah$ dependence in ${\cal N}$.

In an \ads\ background, the fact that $(\eta\l\lh)$ is in the BRST
cohomology allows one to treat $\l^\a$ and $\eta_{\a\ah}\lh^\ah$
as complex conjugates instead of introducing non-minimal variables.
Although the zero mode integral $\int d^{11}\l d^{11}\lh$ diverges
because of the scale factor in $\l$, 
one can easily regularize this divergence by restricting the
zero modes of $\l^\a$ and
$\lh^\ah$ to satisfy $(\eta\l\lh) = \Lambda$ for some positive
constant $\Lambda$. Since $(\eta\l\lh)$ is BRST-invariant, this
regularization preserves BRST invariance. Furthermore, since
the ghost-number anomaly implies that genus $g$ amplitudes
violate ghost-number by $(3-3g,3-3g)$, the dependence on $\Lambda$
can be absorbed by shifting the string coupling constant from
$e^\phi$ to $e^{\phi'}=\Lambda^{-{3\over 2}} e^\phi$.
In other words,
the factor of $e^{(2g-2)\phi'} = \Lambda^{3-3g} e^{(2g-2)\phi}$
at genus $g$ includes the $\Lambda$ dependence.

With this regularization, 
the zero mode integration
for tree amplitudes simplifies to
\eqn\funft{\langle f(x,\t,\l,\th,\lh) \rangle =
\int d^{10}x 
\int d^{16}\t d^{16}\th ~sdet(E_M^A)~ 
\int d^{10}\l d^{10}\lh
~f(x,\t,\l,\th,\lh)}
where $sdet(E_M^A)$ is the superdeterminant of the \ads\
supervierbein and 
$\int d^{10}\l d^{10}\lh$ is an integral over the projective
pure spinors which (after Wick rotation)
parameterize the compact space ${{SO(10)}\over{U(5)}}$.
For example, for three-point supergravity tree amplitudes,
\eqn\fvert{f = (\l^\a\lh^\ah A^{(1)}_{\a\ah}(Z))
(\l^\a\lh^\ah A^{(2)}_{\a\ah}(Z))
(\l^\a\lh^\ah A^{(3)}_{\a\ah}(Z))}
where $\l^\a\lh^\ah A_{\a\ah}(Z)$ is the supergravity vertex operator of \vert.
Integrating over the projective pure spinors
gives
\eqn\treea{\int d^{10}\l\int d^{10}\lh ~f = 
T^{((\a\b\g))((\ah\bh\gh))} A^{(1)}_{\a\ah}(Z) A^{(2)}_{\b\bh} (Z)
A^{(3)}_{\g\gh}(Z)}
where $T^{((\a\b\g))((\ah\bh\gh))}$ is the constant tensor obtained
by symmetrizing $\eta^{\a\ah}\eta^{\b\bh}\eta^{\g\gh}$ with
respect to $(\a\b\g)$ and $(\ah\bh\gh)$ and
removing the gamma-matrix trace terms, i.e. removing the
terms proportional to $\g_m^{\a\b}$ or $\g_m^{\ah\bh}$.  

So the onshell three-point tree amplitude in an \ads\ background
is claimed to be 
\eqn\treebm{
\int d^{10}x 
\int d^{16}\t d^{16}\th ~sdet(E_M^A)~ 
T^{((\a\b\g))((\ah\bh\gh))} A^{(1)}_{\a\ah}(Z) A^{(2)}_{\b\bh}(Z)
A^{(3)}_{\g\gh}(Z).}
It might seem surprising that the zero mode integration in an
\ads\ background selects the term in $(A_{\a\ah})^3$ with 
16 $(\t\th)$'s whereas the zero mode integration in a flat
background selects the term in $(A_{\a\ah})^3$ with 5 $(\t\th)$'s.
However, note that three-point amplitudes in an \ads\ background can
be computed as a sum over $N$-point amplitudes in a flat background where
$(N-3)$ of the vertex operators deform
the flat background to \ads. If 11 of the extra vertex operators
are Ramond-Ramond vertex operators containing the term $\int d^2 z
F^{\a\ah}d_\a \widehat d_\ah$, one could contract 11 $(\t\th)$'s in
$(A_{\a\ah})^3$ with these vertex operators
and convert the flat zero-mode measure factor into
the \ads\ measure factor. So the $\int d^2 z F^{\a\ah}d_\a d_\ah$
term in the \ads\ action of \adsp\ plays the same role as the
$\exp[\rho (\t^\a r_\a +\th^\ah \widehat r_\ah)]$ term in 
the regulator of \regu\ which absorbs 11 $(\t\th)$'s after
integrating over $\int d^{11}r\int d^{11}\widehat r$. 

A separate argument for the validity of the integration measure 
of \funft\ is that it is manifestly $PSU(2,2|4)$ invariant since
it can be written as
\eqn\funfu{\langle f(x,\t,\l,\th,\lh) \rangle =
\int Dg~\int d^{10}\l d^{10}\lh
~f(g,\l,\lh)}
where $g$ is the ${PSU(2,2|4)}\over{SO(4,1)\times SO(5)}$
coset and $Dg$ is the corresponding Haar measure.
For three-point supergravity amplitudes in an \ads\ background,
$PSU(2,2|4)$ invariance together with gauge invariance is
expected to completely
fix the amplitude up to an overall constant. 

This is analogous to
the statement that the three-point supergravity amplitude in
a flat background is completely fixed by super-Poincar\'e invariance
and gauge invariance. In a flat background, the expression
$\int d^{10} x \int d^{16}\t \int d^{16}\th (\l\lh A)^3$ would vanish by
dimensional arguments since it carries 11 too many factors of momentum
and since $k_r\cdot k_s=0$ for on-shell three-point amplitudes.
For this reason, the correct measure factor in a flat background
involves an integration over only 5 $(\t\th)$'s. But
in an \ads\ background, there is no such dimensional argument since
the expression
$\int d^{10} x \int d^{16}\t \int d^{16}\th (\l\lh A)^3$ 
can depend on inverse powers of the $AdS$ radius as $(r_{AdS})^{-11}$.
So assuming that \treebm\ does not vanish for some unknown reason,
$PSU(2,2|4)$ invariance implies that
it must be proportional to the correct three-point supergravity amplitude in
an \ads\ background.

For amplitudes at non-zero genus, the prescription in
the pure spinor formalism
is to insert $(3g-3)$
$b$ and $\bar b$ ghosts and $N$ integrated vertex operators into
the functional integral as in \gloop.
After integrating out the non-zero modes of the worldsheet fields,
one needs to integrate over both the zero modes of $(x,\t,\th,\l,\lh)$
and the $g$ zero modes of the spin-one variables $w_\a$ and $\widehat w_\ah$.
In a flat background, integration over the zero modes of $w_\a$ and $\widehat
w_\ah$ produces divergences which are regularized by including
the term \ntwo\nekrb
\eqn\loopr{\exp[\rho ~Q(\half(\bar\l\g^{ab}s)N_{ab}+
\half(\bar{\lh}\g^{ab}\widehat s)\widehat N_{ab})]}
$$=
\exp [-\rho(N^{ab}\bar N_{ab} + \widehat N^{ab}\bar{\widehat N}_{ab}
-{1\over 4}(\bar\l\g^{ab}s)(\l\g_{ab}d) -{1\over 4}(\bar{\lh}\g^{ab}\widehat s)
(\widehat\l\g_{ab}\widehat d))]$$
in the regulator ${\cal N}$ of \regu\ where
$\bar N_{ab}$ and $\bar{\widehat N}_{ab}$ are the Lorentz currents
for the non-minimal variables and $(s^\a,\widehat s^\ah)$
are the conjugate momenta for $(r_\a,\widehat r_\ah)$.  
However, in an \ads\ background,
the worldsheet action of \adsp\ 
already contains $\exp (-N^{ab}\widehat N_{ab})$
dependence because of the \ads\ curvature which
couples the left and right-moving Lorentz currents. So the curvature
of the \ads\ background acts as a regulator for the $(w_\a,\widehat w_\ah)$
zero mode integration and eliminates the need for the non-minimal
regulator ${\cal N}$ of \loopr.

It should be noted that
because of the non-holomorphic structure of the sigma model, the measure
factor for open string scattering amplitudes in \ads\ will not be
the ``holomorphic square-root'' of the closed string measure factor of \funft.
For example, for $D_3$ branes at the boundary of $AdS_5$, the boundary
condition $\lh^\ah = (\g_{0123})^\ah_\b \l^\b$ implies that
$\l\g^{01234}\lh = \l\g^4\l=0$ because of the pure spinor constraint
$\l\g^a\l=0$. So one cannot impose that $(\eta\l\lh)=0$ on the $D_3$
brane boundary. 

To regularize the functional integral over pure spinors in the presence
of $D_3$ branes, one therefore needs to introduce the same non-minimal
variables $(\bar\l_\a,r_\a)$ on the boundary as one would introduce
in a flat background. After inserting the non-minimal regulator
${\cal N} = \exp [-\rho (\l^\a\bar\l_\a -\t^\a r_\a)]$ on the boundary
and integrating over the non-minimal fields, the zero mode measure factor
for open string amplitudes will involve integration over only 5 $\t$'s.
This is expected since open string amplitudes on \ads\ describe 
${\cal N}=4$ $d=4$ super-Yang-Mills amplitudes which, like $d=10$
super-Yang-Mills amplitudes, are naturally expressed
in pure spinor superspace as integrals over 5 $\t$'s
\ref\superpm
{N. Berkovits, {\it
Covariant quantization of the superparticle using pure spinors},
JHEP 0109 (2001) 016, hep-th/0105050.}\ref\schwarz{M. Movshev and
A.S. Schwarz, {\it On maximally supersymmetric Yang-Mills theories},
Nucl. Phys. B681 (2004) 324, hep-th/0311132.}.

\subsec{Ramond-Ramond plane-wave background}

It is instructive to compare the structure of the zero-mode measure factors
in flat and \ads\ backgrounds with the zero-mode measure factor in
a Ramond-Ramond plane-wave background. The pure spinor action
in this background was described in \rrp\
and has the same structure
as \adsp\ except that the non-vanishing components of $F^{\a\bh}$
and $R_{abcd}$ take the values
\eqn\values{F^{\a\bh} = {1\over{240}} F^{mnpqr} \g_{mnpqr}^{\a\bh} = 
(\g_{-1234})^{\a\bh}, \quad
R_{+j+k} = \d_{jk}}
where $x^{\pm}= x^0 \pm x^9$ and $j=1$ to 8 denote the transverse
directions.

Splitting $d_\a$ and $\widehat d_\ah$ into their $SO(8)$ components as
\eqn\split{d_A = (\g_+\g_- d)_A, 
\quad d_{A'} = (\g_-\g_+ d)_{A'}, \quad
\widehat d_\Ah = (\g_+\g_- \widehat d)_\Ah, 
\quad \widehat d_{\Ah'} = (\g_-\g_+ \widehat d)_{\Ah'} }
where $A,A'=1$ to 8, the term $d_\a F^{\a\bh} \widehat d_\bh$
in \adsp\
implies that $d_A$ and $\widehat d_\Ah$ are auxiliary variables
which can be integrated out. But the variables $d_{A'}$ and
$\widehat d_{\Ah'}$ are propagating and couple to $\t^{A'}=(\g^-\g^+\t)^{A'}$ 
and 
$\th^{\Ah'}=(\g^-\g^+\th)^{\Ah'}$ through the first-order action
\eqn\firsto{\int d^2 z [ d_{A'}\pb\t^{A'} + \widehat d_{\Ah'}\p\th^{A'} ].}

In this plane-wave background, the operator $\eta_{\a\ah} \l^\a \l^\ah$
of \vdil\ is replaced by $\l\g_{+1234}\lh=\eta_{A \Ah} \l^A \lh^\Ah$ where
$\eta_{A \Ah} \equiv (\sigma^{1234})_{A\Ah}$ is constructed from the SO(8)
Pauli matrices $\s^j_{A A'}$ and 
\eqn\splitl{\l^A = (\g^+\g^-\l)^A, \quad \l^{A'}=(\g^-\g^+\l)^{A'},\quad
\lh^\Ah = (\g^+\g^-\lh)^\Ah, \quad \lh^{\Ah'}=(\g^-\g^+\lh)^{\Ah'}.}
Since $\eta_{A\Ah}\l^A\lh^\Ah$ is in the BRST cohomology,
one can treat $\eta_{A\Ah}\lh^\Ah$ as the complex conjugate of
$\l^A$ and impose the constraint that $\eta_{A\Ah}\l^A \lh^\Ah$
is non-vanishing.

This resolves the problem of functional integration over $\l^A$ and $\lh^\Ah$,
but one still needs to regularize the functional integration over the
remaining components $\l^{A'}$ and $\lh^{\Ah'}$ which are $SO(8)$
pure spinors since they satisfy the constraint
\eqn\eightpure{\l^{A'}\l^{A'} = \lh^{\Ah'}\lh^{\Ah'}=0}
coming from the condition $\l\g^+\l = \lh\g^+\lh=0$.
This regularization can be performed by introducing
non-minimal fields $\bar\l_{A'}$ and $\bar{\lh}_{\Ah'}$ and their
BRST superpartners $r_{A'}$ and $\widehat r_{\Ah'}$
which satisfy the constraints
\eqn\eightc{\bar\l_{A'}\bar\l_{A'} = \bar\l_{A'}r_{A'} = 
\bar{\lh}_{\Ah'} \bar{\lh}_{\Ah'} = 
\bar{\lh}_{\Ah'} \widehat r_{\Ah'} = 0.}

One then adds the term $\int dz r_{A'}\bar w^{A'} + \int d\bar z
\widehat r_{\Ah'} \bar{\widehat w}^{\Ah'}$ to the BRST operator
and defines the non-minimal regulator as
\eqn\nonminreg{{\cal N} = \exp [-\rho~Q(\t^{A'}\bar\l_{A'} +\th^{\Ah'}
\bar{\lh}_{\Ah'})] =
 \exp [-\rho (\l^{A'}\bar\l_{A'} -\t^{A'}r_{A'} + 
\lh^{\Ah'}\bar{\lh}_{\Ah'} -\th^{\Ah'}\widehat r_{\Ah'} )]}
Since there are seven independent $r_{A'}$ and $\widehat r_{\Ah'}$
variables, the zero mode integration in a plane-wave background is
of the form
\eqn\fung{\langle f(x,\t,\l,\th,\lh) \rangle =
\int d^{10}x \int d^{11}\l d^{11}\lh d^{7}\bar\l d^{7}\bar{\lh}
\int d^{16}\t d^{16}\th d^{7}r d^{7}\widehat r~{\cal N}~ f(x,\t,\l,\th,\lh)}
$$=\int d^{10}x \int d^8\t^A\int d^8\th^\Ah \int d\l d\lh~~
\int d\t_{A'} {\p\over{\p\l^{A'}}}
~\int d\th_{\Ah'} {\p\over{\p{\lh}^{\Ah'}}}
f(x,\t,\l,\widehat\t,\lh)|_{\t=\th=0}$$
where the integration $\int d\l d\lh$ is over the projective part
of $\l^A$ and $\lh^\Ah$ (keeping $\eta_{A\Ah}\l^A\lh^\Ah$ fixed).
So instead of selecting the term in $f$ with 5 $(\t\th)$'s or 16 $(\t\th)$'s,
the zero mode measure factor in a plane-wave background selects the
term in $f$ with 9 $(\t\th)$'s.

Although this result may seem strange, it is consistent with the
expectation from light-cone gauge analysis. In light-cone gauge,
the supergravity vertex operator in a plane-wave background depends
only on the transverse zero modes and has the form \bmn
\eqn\sugralc{\Phi = f(a_j^\dagger, s_A^\dagger) |0\rangle}
where $a_j^\dagger$ and $s_A^\dagger$ are 8 bosonic and 8 fermionic
operators constructed from the zero modes which ``excite'' the
ground-state wavefunction $|0\rangle$ of the harmonic oscillator 
for the massive zero modes. In terms of the zero modes $(x^j, \t^A,\th^\Ah)$,
the Lagrangian is 
\eqn\lagr{\half \dot x^j \dot x^j + {i\over 2} k^+ (\t^A \dot\t^A
+\th^\Ah \dot{\th}{}^\Ah) - (k^+)^2 (\half x^j x^j + i \eta_{A \Ah}
\t^A \th^\Ah)}
and the ground-state wavefunction is 
\eqn\wavef{|0\rangle = |4\pi k^+|^{-2} \exp(-|k^+| (\half x^j x^j + 
i\eta_{A\Ah}\t^A\th^\Ah))}
where $k^+$ is the $P^+$ momentum of the state.

In light-cone gauge, the measure factor $\langle \Phi_1|\Phi_2\rangle_{LC}$
can be computed either by using the commutation relations of the
operators in \sugralc\ or by evaluating the functional integral
\eqn\functin{\langle\Phi_1|\Phi_2\rangle_{LC} = 
\int d^8 x \int d^8\t\int d^8\th
~\Phi_1 (x^j,\t^A,\th^\Ah)~\Phi_2 (x^j,\t^A,\th^\Ah).}
Note that $|0\rangle $ has a well-defined norm since
\eqn\zerozero
{\langle 0|0\rangle_{LC} = \int d^8 x \int d^8\t \int d^8\th
|4\pi k^+|^{-4} e^{-|k^+| (x^j x^j + 2i\eta_{A\Ah}\t^A\th^\Ah)} = 1.}

The covariant measure factor of \fung\ can be compared with the
light-cone measure factor of \functin\ using the relation that
$\langle  V_1 | c_0 \bar c_0 | V_2 \rangle $ should be proportional
to 
$\langle \Phi_1 |\Phi_2\rangle_{LC}$
where $V$ is the BRST-invariant vertex operator of ghost-number
$(1,1)$ corresponding to the light-cone vertex operator $\Phi$, 
and
$c_0$ and $\bar c_0$ are operators satisfying $\{b_0,c_0\} =
\{\bar b_0,\bar c_0\}=1$. 
The factors of $c_0$ and $\bar c_0$ come from BRST gauge-fixing and
are necessary for the covariant measure factor to have ghost-number
$(3,3)$.

In a plane-wave background, the BRST-invariant vertex operator
corresponding to the light-cone field $\Phi(x^j,\t^A,\th^\Ah)$ is
\eqn\covver{V = \l^\a \lh^\ah A_{\a\ah}(x,\t,\th) = 
(\eta_{A\Ah}\l^A\lh^\Ah) \Phi(x^j,\t^A,\th^\Ah) e^{ik^+ x^- + ik^- x^+}
+ ...}
where $\Phi$ is the light-cone superfield of \sugralc\ and
$...$ depends on $\t^{A'}$ and $\th^{\Ah'}$ and is determined
by BRST invariance. Furthermore, since the $b$ and $\bar b$ ghosts in the pure
spinor formalism have the term
\eqn\bghs{b = (\bar\l_{A'}\l^{A'})^{-1} \p x^+ (\bar\l\g^- d) + ...,\quad
\bar b = (\bar\l_{A'}\l^{A'})^{-1} \bar\p x^+ (\bar{\lh}\g^- \widehat
d) + ...,}
one can define $c_0$ and $\bar c_0$ satisfying
$\{b_0,c_0\} = \{\bar b_0,\bar c_0\} =1$ as 
\eqn\ccbar{c_0 = [(\p x^+)^{-1} \l^{A'} \t^{A'} ]_0
= (k^+)^{-1} \l^{A'}\t^{A'},\quad\bar c_0 = 
[(\bar\p x^+)^{-1} \lh^{A'} \th^{A'} ]_0 =
(k^+)^{-1} \lh^{\Ah'}\th^{\Ah'}.}

So the covariant measure factor of 
\fung\ implies that
\eqn\covtwo{\langle  V_1 | c_0 \bar c_0 | V_2 \rangle =
\int d^{10}x \int d^8\t^A\int d^8\th^\Ah \int d\l d\lh~~
\int d\t_{A'} {\p\over{\p\l^{A'}}}
~\int d\th_{\Ah'} {\p\over{\p{\lh}^{\Ah'}}}} 
$$(\eta_{A\Ah}\l^A\lh^\Ah)^2 ~
\Phi_1 \Phi_2 ~(k^+)^{-2} (\l^{A'}\t^{A'})(\lh^{\Ah'}\th^{\Ah'})
e^{i(k^+_1+k^+_2)x^- + i(k^-_1+k^-_2)x^+}$$
$$ =(k^+)^{-2}
\d(k^+_1+k^+_2)\d(k^-_1+k^-_2)\int d^8 x\int d^8\t^A\int d^8\th^\Ah
~\Phi_1\Phi_2,$$
which is proportional to the light-cone measure factor 
$\langle \Phi_1|\Phi_2\rangle_{LC}$ of \functin.

So in a plane-wave background,
the covariant measure factor involving integration over 9 $(\t\th)$'s
is related to light-cone integration over 8 $(\t\th)$'s plus
an additional integration over $\t\th$ coming from the $c_0\bar c_0$ term.
In a flat background, the covariant measure factor of \funf\
involving integration
over 5 $(\t\th)$'s can be similarly related to light-cone integration
over 4 $(\t\th)$'s plus an integration over $\t\th$ coming from the
$c_0\bar c_0$ term. In light-cone gauge
in a flat background, the fermionic zero modes
are massless and in order to construct normalizable wavefunctions,
the $SO(8)$ components $\t^A$ and $\th^A$ need
to be split into $U(4)$ components as $(\t^I,\bar\t_I)$ and 
$(\th^{\widehat I},\bar\th_{\widehat I})$ for $I,\widehat I=1$ to 4
\ref\brink{L. Brink, O. Lindgren and B.E.W. Nilsson, {\it N=4 Yang-Mills
on the light cone}, Nucl. Phys. B212 (1983) 401.}.
The resulting light-cone wavefunction is a chiral superfield
$\Phi(\t^I,\th^{\widehat I})$ satisfying the reality condition
\eqn\reai{D_I D_J \widehat D_{\widehat I}\widehat D_{\widehat J} \Phi =
{1\over 4}\epsilon_{IJKL}\epsilon_{\widehat I\widehat J\widehat K\widehat L}
\bar D^K \bar D^L \bar{\widehat D}^{\widehat K}\bar{\widehat D}^{\widehat L} 
\bar\Phi, }
and the light-cone measure factor in a flat background is
\eqn\flatzm{\langle\Phi_1|\Phi_2\rangle_{LC}= 
\int d^8 x\int d^4\t^I\int d^4\th^{\widehat I} ~\Phi_1 \Phi_2}
which involves an integration over only 4 $(\t\th)$'s.

\newsec{Topological $AdS_5\times S^5$ Sigma Model}

In this section, a BRST-trivial action will be constructed with
the same BRST operator and stress-tensor as the \ads\ action of \class, 
and
will be shown to arise from gauge-fixing the ${ G}/{ G}$
principal chiral model where ${ G}=PSU(2,2|4)$. This topological
action will then be argued to describe the zero-radius limit of \ads\
by comparing its physical states with the spectrum of 
gauge-invariant operators of
free ${\cal N}=4$ $d=4$ super-Yang-Mills. A handwaving argument
based on open-closed topological duality 
will then be proposed for showing that the scattering amplitudes
of this topological string coincide with super-Yang-Mills scattering
amplitudes in the limit of small `t Hooft coupling constant.

\subsec{Topological action}

Because of the possibility of including $(\eta\l\lh)^{-1}$ dependence
in the action, one can construct a BRST-trivial action which has the
same stress tensor as the \ads\ action of \class. 
This topological action is
\eqn\topa{S_{top} = \int d^2 z ~Q(\Psi)}
$$= \int d^2 z [{{ \eta^{\a\ah} (\g_a\l)_\a (\g_b\lh)_\ah}\over{2(\eta\l\lh)}}
J^a \bar J^b + \eta_{\a\ah}\bar J^\a J^\ah - w_\a \bar\nabla \l^\a
+ \ww_\ah \nabla \lh^\ah -\eta_{[ab][cd]}N^{ab}\widehat N^{cd} +
\eta^{\a\ah} w_\a^*\ww_\ah^*]$$
where
\eqn\gaugeom{\Psi = \half (\eta\l\lh)^{-1} \lh^\ah
(\half \g_{a\ah\bh} \bar J^a J^\bh +{1\over 4} (\g_{ab})_\ah^\bh \eta_{\b\bh}
N^{ab} \bar J^\b +{1\over 4} \eta_{\a\ah} J_{gh} \bar J^\a)}
$$+\half (\eta\l\lh)^{-1} \l^\a
(-\half \g_{a\a\b} J^a \bar J^\b -{1\over 4} (\g_{ab})_\a^\b \eta_{\b\bh}
\widehat N^{ab} J^\bh -{1\over 4} \eta_{\a\ah} \widehat J_{gh} J^\ah)$$
$$
+\half \eta^{\a\ah} (w_\a \ww_\ah^* -w_\a^* \ww_\ah).$$
Note the close resemblence of the first two
lines in $\Psi$ with the $b$ and $\bar b$
ghost of \bnew\ and \defbarb, 
and that the last line of $\Psi$ is gauge-invariant under \wwgauge\ 
because of the constraints of \constrw. 
Since $Q$ is nilpotent, \topa\ is invariant 
under the BRST transformation
of \brstt\ and \modqw\ and the resulting Noether charge is
\eqn\noeth{Q = \int dz \eta_{\a\ah}\l^\a J^\ah + \int d\bar z
\eta_{\a\ah}\lh^\ah \bar J^\a}
as before.

Using the identity of \identt\ and the
BRST transformations of \brstt\ and \modqw, 
it is straightforward to show that
$Q\Psi$ is equal to the Lagrangian of \topa.
The BRST transformation of the first line of \gaugeom\ is 
\eqn\firstline{
\half[{{ \eta^{\a\ah} (\g_a\lh)_\ah (\g_b\l)_\a}\over{2(\eta\l\lh)}}
\bar J^a J^b + \eta_{\a\ah}\bar J^\a J^\ah - w_\a \bar\nabla \l^\a
+{1\over {8(\eta\l\lh)}} ((w^*\g^{ab} \l)(\lh\g_{ab}\bar J) + 2(w^*\l)
(\lh\bar J))],}
the BRST transformation of the second line of \gaugeom\ is 
\eqn\secondline{
\half[{{ \eta^{\a\ah} (\g_a\l)_\a (\g_b\lh)_\ah}\over{2(\eta\l\lh)}}
J^a \bar J^b + \eta_{\a\ah}\bar J^\a J^\ah + \ww_\ah \nabla \lh^\ah
-{1\over {8(\eta\l\lh)}} ((\ww^*\g^{ab} \lh)(\l\g_{ab} J) + 2(\ww^*\lh)
(\l J))],}
and the BRST transformation of the third line of \gaugeom\ is 
\eqn\thirdline{
\half[2\eta^{\a\ah} w^*_\a \ww^*_\ah + w_\a^* \bar J^\a - \ww_\ah^* J^\ah
- w_\a \bar\nabla \l^\a + \ww_\ah \nabla \lh^\ah 
- 2 \eta_{[ab][cd]}N^{ab}\widehat N^{cd}].}

It is interesting to note that the difference between the topological
and \ads\ actions of \topa\ and \class\ is
\eqn\diffs{S_{top}-S_{AdS_5\times S^5} = \int d^2 z[ 
{{ \eta^{\a\ah} (\g_a\l)_\a (\g_b\lh)_\ah}\over{4(\eta\l\lh)}}
(J^a \bar J^b - \bar J^a J^b) + {1\over 4} \eta_{\a\bh}
(J^\a \bar J^\bh - \bar J^\a J^\bh)],}
where the pure spinors $(\l^\a,\lh^\ah)$ choose a complex structure
which allows the covariant construction of a Wess-Zumino term
from the bosonic currents $(J^a,\bar J^a)$. Using $\l\g^a\l = \lh\g^a\lh=0$
and the BRST transformation of \brstt, 
one can easily check that \diffs\ is
BRST-closed. And since \diffs\ is antisymmetric
in $z$ and $\bar z$, it is clear that the stress tensor of $S_{top}$
is equal to the \ads\ stress tensor of \adsstr.

One can formally define an analogous topological action in a flat Type II
background as 
\eqn\topflat{S^{flat}_{top} = \int d^2 z ~Q(\Psi^{flat})}
$$= \int d^2 z [{{ \eta^{\a\ah} (\g_a\l)_\a (\g_b\lh)_\ah}\over{2(\eta\l\lh)}}
\Pi^a \bar \Pi^b - d_\a \bar\p\t^\a + \widehat d_\ah \p\th^\ah
-w_\a \bar \p \l^\a + \ww_\ah\p\lh^\ah 
+\eta^{\a\ah} w_\a^*\ww_\ah^*]$$
where $\Pi^a = \p x^a + \t\g^a\p\t + \th\g^a\p\th$, $\eta^{\a\ah}$ is
a constant bispinor, and 
\eqn\gaugeom{\Psi^{flat} = \half (\eta\l\lh)^{-1} \lh^\ah \eta_{\a\ah}
(\half \g_a^{\a\b} \bar \Pi^a d_\b 
+{1\over 4} (\g_{ab})^\a_\b 
N^{ab} \bar\p \t^\b +{1\over 4} J_{gh} \bar \p\t^\a)}
$$+\half (\eta\l\lh)^{-1} \l^\a \eta_{\a\ah}
(-\half \g_a^{\ah\bh} \Pi^a \widehat d_\bh -{1\over 4} 
(\g_{ab})_\bh^\ah 
\widehat N^{ab} \p\th^\bh -{1\over 4} \widehat J_{gh} \p\th^\ah) $$
$$ +\half \eta^{\a\ah} (w_\a \ww_\ah^* -w_\a^* \ww_\ah).$$

The choice of $\eta^{\a\ah}$ breaks Lorentz invariance for the Type IIB
superstring, but for the Type IIA superstring, Lorentz invariance
can be preserved by choosing $\eta^{\a\ah}=\d^{\a\ah}$.
Note that unlike the usual pure spinor action in a flat background, 
the topological action $S^{flat}_{top}$ is manifestly spacetime supersymmetric
and satisfies
\eqn\diffflat{
S^{flat}_{top}-S_{flat} = \int d^2 z[ 
{{ \eta^{\a\ah} (\g_a\l)_\a (\g_b\lh)_\ah}\over{4(\eta\l\lh)}}
(\Pi^a \bar \Pi^b - \bar \Pi^a \Pi^b) - L_{WZ}]}
where $L_{WZ}$ is the standard Green-Schwarz Wess-Zumino term.
However, unlike the topological \ads\ action of \topa, the topological
action of \topflat\ in a flat background is not well-defined since inverse
powers of $(\eta\l\lh)$ are not allowed in the flat Hilbert space.
As emphasized in section 3, the presence of inverse powers of $(\eta\l\lh)$
in a flat background would trivialize the BRST cohomology.

\subsec{{ G}/{ G} principal chiral model}

In \topone\
and \limtwo,
an $A$-twisted $N=2$ worldsheet supersymmetric sigma model
constructed from the fermionic coset ${{PSU(2,2|4)}\over{SO(4,2)\times 
SO(6)}}$ was conjectured to describe the zero-radius limit of the \ads\
superstring. This topological sigma model was related by a field
redefinition to the \ads\ sigma model of \class, but the BRST operators
for the topological and \ads\ sigma models were different. It was then
shown in \topthree\
that this 
$N=2$ worldsheet supersymmetric sigma model
constructed from the fermionic coset ${{PSU(2,2|4)}\over{SO(4,2)\times 
SO(6)}}$ could be obtained by gauge-fixing the ${ G}/{ G}$
principal chiral model
\eqn\chiral{S = Str \int d^2 z (G^{-1}\p G - A)(G^{-1}\bar\p G - \bar A)
=
\int d^2 z ~\eta_{\tilde A\tilde B}~(J^{\tilde A}- A^{\tilde A})
(\bar J^{\tilde B} - \bar A^{\tilde B})}
where $G$ takes values in $PSU(2,2|4)$, $J= G^{-1}\p G$ are the left-invariant
currents, 
$\eta_{\tilde A\tilde B}$ is the $PSU(2,2|4)$ metric, and 
$(A,\bar A)$ is a worldsheet gauge field taking values in the
$PSU(2,2|4)$ Lie algebra. Although this ${ G}/{ G}$
model appears to be trivial, it will be argued later that it contains
non-trivial physical states because of boundary conditions on the non-compact
$PSU(2,2|4)$ generators.

The action of \chiral\ is invariant under the local $PSU(2,2|4)$
gauge transformations
\eqn\localpsu{\d G = G\Omega,\quad \d A = d\Omega + [A,\Omega],}
and to obtain the supersymmetric
sigma model based on the fermionic coset, one first uses the
$SO(4,2)\times SO(6)$ generators of $\Omega$ to gauge away
the bosonic elements in $G$ so that $G$ takes values in
the fermionic coset 
${{PSU(2,2|4)}\over{SO(4,2)\times 
SO(6)}}$.
One then uses the fermionic generators of $\Omega$ to gauge-fix
\eqn\onegauge{A^{\a +} \equiv A^\a +i A^\ah =0, \quad 
\bar A^{\a -} \equiv \bar A^\a -i \bar A^\ah =0, }
where $T_{\a +} \equiv T_\a +i T_\ah$ are the 16 fermionic generators
in the upper-right square of $PSU(2,2|4)$ and 
$T_{\a -} \equiv T_\a -i T_\ah$ are the 16 fermionic generators
in the lower-left square of $PSU(2,2|4)$. 

This fermionic gauge-fixing
gives rise to bosonic ghosts $(Z^{\a -},\bar Z^{\a +})$ and
antighosts $(Y_{\a -},\bar Y_{\a +})$ with the Faddeev-Popov action
\eqn\sghosto{S_{gh} = \int d^2 z [ -Y_{\a -} \bar\nabla Z^{\a -}
+ \bar Y_{\a +}\nabla \bar Z^{\a +}]}
and the BRST operator
\eqn\brstone{Q = \int dz \eta_{\a \b} Z^{\a -} J^{\b +}
+\int d\bar z \eta_{\a \b}\bar Z^{\b +} \bar J^{\a -}}
where $\eta_{\a \b } = (\g^{01234})_{\a\b}$. 
Note that $Q^2=0$ without imposing pure spinor constraints on
$Z^{\a -}$ and $\bar Z^{\a +}$ because $T_{\a +}$ and $T_{\a -}$
satisfy $\{T_{\a +}, T_{\b +}\}=
\{T_{\a -}, T_{\b -}\}= 0$.
In this gauge, the action of \chiral\ reduces to an $A$-twisted 
$N=2$ worldsheet supersymmetric sigma model where 
$(Z^{\a -},\bar Z^{\a +},Y_{\a -},\bar Y_{\a +})$ are the
bosonic worldsheet superpartners to the fermionic coset 
${{PSU(2,2|4)}\over{SO(4,2)\times 
SO(6)}}$ and \brstone\ is the scalar worldsheet supersymmetry generator.

Although the BRST operator of \brstone\ in this gauge-fixing
is different from the original
\ads\ BRST operator of \brstp, it will now be shown that there is an alternative
gauge-fixing of the ${ G}/{ G}$ model of \chiral\ which leads
to the topological action of \topa\
and which has the same BRST operator as \brstp.
To obtain the topological action of \topa\ from \chiral, 
one first uses the local
$SO(4,1)\times SO(5)$ gauge invariances of \localpsu\ to gauge-fix $G$ to
take values in the Metsaev-Tseytlin coset ${{PSU(2,2|4)}\over{SO(4,1)\times
SO(5)}}$. One next uses the fermionic gauge transformations of \localpsu\ to
gauge-fix
\eqn\gaugetwo{A^\ah =0,\quad \bar A^\a =0,}
which gives rise to unconstrained bosonic ghosts
$(Z^\a,\bar Z^\ah)$ and antighosts $(Y_\a,\bar Y_\ah)$ with the
Faddeev-Popov action
\eqn\sghostt{S_{gh} = \int d^2 z [ -Y_\a \bar\nabla Z^\a
+ \bar Y_\ah\nabla \bar Z^\ah]}
where $\bar\nabla Z^\a = \bar\p Z^\a + \half \bar A^{[ab]} (\g_{[ab]} Z)^\a$
and $\nabla \bar Z^\ah = \p \bar Z^\ah + \half A^{[ab]} (\g_{[ab]}\bar Z)^\ah$.
Since $\{T_\a,T_\b\}$ and 
$\{T_\ah,T_\bh\}$ are nonzero and $Z^\a$ and $\bar Z^\ah$
are unconstrained, the BRST operator
\eqn\brstche{Q = \int dz \eta_{\a\ah} Z^\a J^\ah + \int d\bar z
\eta_{\a\ah} \bar Z^\ah \bar J^\a}
implied by this gauge-fixing would not be nilpotent.

However, one still has ten bosonic gauge transformations of \localpsu\ which
need to be gauge-fixed. Although one could naively use these
gauge transformations to gauge away the remaining bosonic components
of $G$, this will be argued later to be inconsistent with the boundary
conditions of the $PSU(2,2|4)$ gauge parameters. Instead, one
can use these ten gauge transformations to gauge-fix 5 components of
$A^a$ and 5 components of $\bar A^a$ to zero.
The choice of which five components of $A^a$ and $\bar A^a$ are
gauge-fixed will be correlated with the bosonic ghosts $(Z^\a,\bar Z^\ah)$
in such a manner that the resulting BRST operator is nilpotent.
Using an \ads\ adaptation of the ``extended pure spinor formalism'' of
Aisaka and Kazama \aisaka, this BRST operator will then be shown to have the
same cohomology as the original \ads\ BRST operator of \brstp.

To determine which components of $A^a$ should be gauge-fixed,
note that $(\g_a)_{\a\b} Z^\a Z^\b$ is a null vector which decomposes under
$SO(4,1)\times SO(5)$ into
\eqn\decompz{\Phi_I = (\g_I)_{\a\b} Z^\a Z^\b,
\quad \Psi_{\tilde I} = (\g_{\tilde I})_{\a\b} Z^\a Z^\b}
for $I=0$ to 4 and $\tilde I=5$ to 9.
Furthermore, if
$\Phi_I$ is zero for $I=0$ to 4, then  
$\Psi_{\tilde I}$ is also zero for $\tilde I=5$ to 9.
This can be seen from the fact that a pure spinor contains
11 independent components and therefore satisfies 5 independent
constraints. So if $\Phi_I=0$ for $I=0$ to 4, $Z^\a$ will
be a pure spinor, which implies that 
$\Psi_{\tilde I}=0$ for $\tilde I=5$ to 9.
Since $\Phi_I=0$ implies $\Psi_{\tilde I}=0$, 
there exists an invertible matrix $M^J_{\tilde I}(Z)$ such that
\eqn\invrel{\Psi_{\tilde I}(Z) = M^J_{\tilde I}(Z) ~\Phi_J(Z).}

It will be convenient to define the matrix ${\cal N}_a^I(Z)$ such that
\eqn\Ndef{\g_{a\a\b} Z^\a Z^\b = {\cal N}_a^I(Z) \Phi_I(Z)}
where ${\cal N}_a^I = \d_a^I$ for $a=0$ to 4, and ${\cal N}_a^I = M_a^I$ for
$a=5$ to 9. Since $\eta^{ab} (Z\g_a Z)(Z\g_b Z)=0$ and since
the $\Phi_I$'s are independent, ${\cal N}_a^I$ satisfies the identity
\eqn\Nident{\eta^{ab} {\cal N}_a^I {\cal N}_b^J = 0.} Similarly, one can
define the matrix $\bar {\cal N}_a^I(\bar Z)$ such that
\eqn\Nbarident{\g_{a\ah\bh} \bar Z^\ah \bar Z^\bh = \bar {\cal N}_a^I(\bar Z)
\bar\Phi_I(\bar Z), \quad \eta^{ab} \bar {\cal N}_a^I \bar {\cal N}_b^J=0.}

One now uses ${\cal N}_a^I(Z)$ and $\bar {\cal N}_a^I(\bar Z)$ to choose the
gauge-fixing conditions
\eqn\gaugefixc{ {\cal N}_a^I(Z) A^a =0,\quad \bar {\cal N}_a^I(\bar Z) \bar A^a =0 }
for $I=0$ to 4. With this gauge-fixing, the ${ G}/{ G}$ model
of \chiral\ becomes
\eqn\actchir{S = \int d^2 z [\eta_{\tilde A\tilde B} (J^{\tilde A} -
A^{\tilde A})(\bar J^{\tilde B}-\bar A^{\tilde B}) 
+ \bar f_I {\cal N}^I_a A^a + f_I \bar {\cal N}^I_a \bar A^a  + f_\a \bar A^\a +
\bar f_\ah A^\ah}
$$-Y_\a (\bar\nabla Z^\a -\eta^{\a\ah}(\bar Z\g_a)_\ah \bar A^a
-c^a \g_a^{\a\b} \eta_{\b\bh} \bar A^\bh)
+ \bar Y_\ah (\nabla \bar Z^\ah 
+\eta^{\a\ah}(Z\g_a)_\a  A^a
+c^a \g_a^{\ah\bh} \eta_{\b\bh} A^\b)$$
$$-b_I \bar {\cal N}^I_a (
\bar\nabla c^a +(\bar Z\g^a)_\ah \bar A^\ah + (Z\g^a)_\a \bar A^\a)
- \bar b_I {\cal N}^I_a(\nabla c^a
+(Z\g^a)_\a  A^\a + (\bar Z\g^a)_\ah A^\ah)
]$$
and the BRST operator is 
\eqn\brstone{Q = \int dz [Z^\a f_\a + b_I R^{IJ} \Phi_J + c^a 
(\bar {\cal N}_a^I f_I +
K_a)]}
$$ +\int d\bar z
[\bar Z^\ah \bar f_\ah  + \bar b_J R^{IJ} \bar\Phi_I + c^a 
({\cal N}_a^I \bar f_I +
\bar K_a)] $$
where $(f_I,\bar f_I, f_\a,\bar f_\ah)$ are Lagrange multipliers
which impose the gauge-fixing conditions, $(c^a, Z^\a, \bar Z^\ah)$ and
$(b_I, \bar b_I, Y_\a, \bar Y_\ah)$ are the Faddeev-Popov ghosts and
antighosts coming from the gauge-fixing of \gaugetwo\ and \gaugefixc, and
\eqn\rdeff{R^{IJ} \equiv \eta^{ab} \bar{\cal N}_a^I {\cal N}_b^J, \quad
K_a \equiv \eta_{\a\ah} (\g_a Y)^\a \bar Z^\ah, \quad
\bar K_a \equiv \eta_{\a\ah} (\g_a \bar Y)^\ah  Z^\a.}

After integrating out the worldsheet gauge fields and Lagrange multipliers
which satisfy auxiliary equations of motion, \actchir\ reduces to
\eqn\actred{S = \int d^2 z [ J^a {\cal N}_a^I R^{-1}_{IJ} \bar {\cal N}_b^J \bar J^b
+ \eta_{\a\ah} \bar J^\a J^\ah -
Y_\a (\bar\nabla Z^\a  + ... ) + \bar Y_\ah (\nabla \bar Z^\ah  + ...)}
$$
-b_I \bar {\cal N}^I_a (\bar\nabla c^a + ...)
 +\bar b_I {\cal N}^I_a (\nabla c^a + ...)
- \eta^{[ab][cd]}(\half Y\g_{ab}Z + b_I \bar {\cal N}^I_a c_b)(\half
\bar Y\g_{cd}\bar Z + \bar b_J {\cal N}^J_c c_d) ]
$$
with the BRST operator
\eqn\brstnewt{
Q = \int dz [\eta_{\a\ah} Z^\a J^\ah + 
b_I R^{IJ} \Phi_J + c^a \bar {\cal N}_a^I R_{JI}^{-1} {\cal N}_b^J (J^b - K^b) + c^a K_a] }
$$+\int d\bar z
[\eta_{\a\ah} \bar Z^\ah \bar J^\a + \bar b_I R^{JI} \bar\Phi_J + 
c^a {\cal N}_a^I R_{IJ}^{-1} \bar {\cal N}_b^J (\bar J^b - \bar K^b) + c^a \bar K_a] $$
where 
\eqn\defnn{\bar\nabla Z^\a = 
\bar\p Z^\a +\half \bar J^{[ab]}(\g_{ab} Z)^\a,\quad
\nabla \bar Z^\ah = \p \bar Z^\ah +\half J^{[ab]}(\g_{ab} \bar Z)^\ah,}
$$\bar\nabla c^a = \bar\p c^a + \bar J^{[ab]} c_b, \quad
\nabla c^a = \p  c^a + J^{[ab]} c_b, $$
and $R^{-1}_{IJ}$ is the inverse matrix to 
$R^{IJ}\equiv\eta^{ab} \bar {\cal N}_a^I {\cal N}_b^J$ satisfying
$R^{-1}_{IJ} R^{JK} = \d_I^K$.
Note that the last term of \actred\ comes from integrating out 
$A^{[ab]}$ and $\bar A^{[ab]}$ which converts the covariant
derivatives in \sghostt\ into the covariant derivatives of \defnn.

As shown in \aisaka\ using 
``homological perturbation'' theory, the BRST operator of
\brstnewt\ is equivalent to the BRST operator
$Q=\int dz \eta_{\a\ah}\l^\a J^\ah + \int d\bar z\eta_{\a\ah} 
\lh^\ah \bar J^\a$
where the terms $\int dz b_I R^{IJ}\Phi_J$ and
$\int d\bar z \bar b_I R^{JI}\bar\Phi_J$ in \brstnewt\ have been
used to strongly impose the constraints $\Phi_I = \bar \Phi_I=0$
and to gauge $c^a=0$. In the presence of the constraints $\Phi_I=\bar\Phi_I=0$,
the ghosts $Z^\a$ and $\bar Z^\ah$ reduce to pure spinors which will
be called $\l^\a$ and $\lh^\ah$. Furthermore, $\Phi_I=\bar\Phi_I=0$
implies that 
$(\l\g^a)_\a {\cal N}_a^I=
(\lh\g^a)_\ah \bar{\cal N}_a^I= 0$, and that
\eqn\ipli{{\cal N}_a^I R^{-1}_{IJ} \bar {\cal N}_b^J =
{{(\l \g_a)_\a \eta^{\a\ah} (\lh\g_b)_\ah }\over{2(\eta\l\lh)}}}
where the normalization of \ipli\ is fixed by
$\eta^{ab}
({\cal N}_a^I R^{-1}_{IJ} \bar {\cal N}_b^J)=R_{IJ}^{-1} R^{JI}=5$. 
Finally, when $c^a=0$ and $\Phi_I=\bar \Phi_I=0$, it is straightforward
to check that the $...$ terms in \actred\ are zero and that \actred\ coincides
with \topa.

So it has been shown that the topological \ads\ action of \topa\
and BRST operator
of \noeth\ can be obtained from the ${ G}/{ G}$ principal
chiral model of \chiral\ by choosing the gauge 
\eqn\gaugechi{A^\ah = \bar A^\a = {\cal N}^I_a(Z) A^a = \bar {\cal N}^I_a(\bar Z)
\bar A^a =0,}
where the tensors ${\cal N}_a^I(Z)$ and $\bar {\cal N}_a^I(\bar Z)$ are constructed
from the bosonic Faddeev-Popov ghosts.
In the next subsection, it will be argued that this topological model
describes the zero-radius limit of the \ads\ superstring which is
dual to free ${\cal N}=4$ $d=4$ super-Yang-Mills theory.

\subsec{ Physical states}

If the topological model of \topa\ is to describe the zero radius limit
of the \ads\ superstring, physical states in the BRST cohomology of
this model should correspond to gauge-invariant super-Yang-Mills
operators at zero `t Hooft coupling.
Naively, the ${ G}/{ G}$ model has 
no physical states since one could use the local $PSU(2,2|4)$
gauge invariance of \localpsu\ to gauge $G=1$. In this gauge, there are no
propagating ghosts and the equations of motion for the worldsheet
gauge field are simply $A^{\tilde A} = \bar A^{\tilde A}=0$.

However, because of the non-compact generators in $PSU(2,2|4)$,
there are subtleties in choosing the gauge $G=1$. Suppose one
parameterizes the $PSU(2,2|4)$ matrix $G$ as
\eqn\paramg{G = \exp (x^m P_m + \t^j_\m q^\m_j + \tb^\md_j \bar q_\md^j)
\exp (- y D + \phi_{jk} R^{jk} + t_{mn} M^{mn}) \exp (h_m K^m 
+ \xi^\m_j s_\m^j + \bar\xi_\md^j \bar s^\md_j)}
where $(P_m, q^\m_j, \bar q_\md^j)$ are the ${\cal N}=4$ $d=4$
translation and supersymmetry generators for $m=0$ to 3, $j=1$ to 4
and $(\m,\md)=1$ to 2, $(D, R^{jk}, M^{mn})$ are the
dilatation, $SO(6)$ $R$-symmetry, and $SO(3,1)$ Lorentz generators, 
and $(K^m, s_\m^j, \bar s^\md_j)$ are the conformal and superconformal
generators. With this parameterization of $G$, the global $PSU(2,2|4)$
isometries $\d G = \Sigma G$ transform the variables $(x^m,\t^j_\m,\tb^\md_j)$
into themselves in the standard ${\cal N}=4$ $d=4$ superconformal manner.
Furthermore, using the relations
\eqn\relationone{K^m e^{-y D} = e^{-yD} (e^{-y} K_m), \quad
s_\m^j e^{-y D} = e^{-yD} (e^{-\half y} s_\m^j), \quad
\bar s^\md_j e^{-y D} = e^{-yD} (e^{-\half y} \bar s^\md_j), }
one finds that in the limit $y\to \infty$, the variables
$(h^m,\xi_j^\m,\bar\xi_\md^j)$ are invariant under the global $PSU(2,2|4)$
transformations.
So it is natural to identify $(x^m,\t^j_\m,\tb^\md_j)$ as parameterizing
the boundary of \ads\ in the limit where $y\to \infty$.

Under the local $PSU(2,2|4)$ gauge transformations $\d G= G\Omega$ of 
\localpsu,
one could naively gauge-fix to zero all the variables in \paramg.
However, using the relations
\eqn\relationtwo{ e^{-y D} P_m = (e^{-y} P_m) e^{-yD} , \quad
e^{-y D} q^\m_j =  (e^{-\half y} q^\m_j) e^{-yD}, \quad
e^{-y D} \bar q_\md^j =  (e^{-\half y} \bar q_\md^j) e^{-y D}, }
one finds that in the limit where $y\to \infty$, the variables
$(x^m,\t^j_\m,\tb^\md_j)$ are invariant under these gauge transformations.
So assuming that the gauge parameters in $\Omega$ of \localpsu\ do not blow up
when $y\to\infty$, the boundary of \ads\ is gauge-invariant and
cannot be gauged away. The ${ G}/{ G}$ principal chiral model
could therefore have physical states which depend non-trivially
on the \ads\ boundary variables 
$(x^m,\t^j_\m,\tb^\md_j)$ when $y\to\infty$.\foot{
Using the gauge-fixing to the fermionic
coset, the $x^m$ variables
were gauged to zero which explains why it was difficult to
construct physical vertex operators in terms of the fermionic
coset variables.
In \topthree, it was conjectured that the non-trivial physical
states could emerge after including a kinetic term 
for the worldsheet gauge field. However,
this conjecture appears to be incorrect
since the kinetic term goes to zero in the infrared 
limit of the sigma model. I would like to thank A. 
Polyakov for correcting this point and for suggesting that the
topological action should be perturbed by an appropriate
radius-dependent operator.}

In fact, it is easy to verify that in the gauge of \gaugechi\ where
$G$ takes values in the Metsaev-Tseytlin coset $g\in {{PSU(2,2|4)}\over
{SO(4,1)\times SO(5)}}$, 
there are such physical states
in the BRST cohomology.
Using the topological action of \topa, the BRST
operator of \noeth\ transforms 
\eqn\brstaga{Qg = g (\l^\a T_\a + \lh^\ah T_\ah)}
in precisely the same manner as in the \ads\ formalism of section 2.
So the supergravity vertex operator $V=\l^\a \lh^\ah A_{\a\ah}(x,\t,\th)$
is in the BRST cohomology of the topological model when $A_{\a\ah}$
satisfies the equations of motion and gauge invariances of \eqA\
and \agauge.

These supergravity vertex operators depend only on the zero modes of
the worldsheet variables and correspond to the half-BPS Yang-Mills
operators. Vertex operators corresponding to non-BPS Yang-Mills operators
are expected to depend on non-zero modes of the worldsheet variables
and will be more difficult to explicitly construct. Nevertheless, it
will be conjectureed that these non-BPS vertex operators can be obtained
from BPS vertex operators by transforming the worldsheet variables
described by the Metsaev-Tseytlin
coset $g\in {{PSU(2,2|4)}\over{SO(4,1)\times SO(5)}}$
as 
\eqn\conjt{\d g (\s) =  \Sigma(\s) g(\s)}
where $0\leq\s < 2\pi$ is the closed string parameter and
$\Sigma(\s)$ is a $PSU(2,2|4)$ transformation which is allowed to
depend on $\s$.

Since \conjt\ acts by left multiplication and the BRST transformation
of \brstaga\ acts by right multiplication,
BRST transformations commute with \conjt. So $Q V(g)=0$ implies
that $Q V(g + \d g) =0$ where $\d g$ is defined in \conjt.
When $\Sigma$ is independent of $\s$, \conjt\ is a global $PSU(2,2|4)$
transformation which takes half-BPS vertex operators into half-BPS
vertex operators. But when $\Sigma$ depends on $\s$, \conjt\ can
take half-BPS vertex operators into non-BPS vertex operators which
depend on non-zero modes of the worldsheet variables. Although \conjt\
does not leave invariant the topological action of \topa\ when $\p_\s \Sigma$
is nonzero, the change of the topological action is BRST-trivial
and can be expressed as $\d S = \int d^2 z Q [\Psi(g + \d g) - \Psi(g)]$
where $\Psi$ is defined in \gaugeom. So the transformation of \conjt\ takes
physical states into physical states.

To see an example where
\conjt\ transforms a physical half-BPS vertex operator into
a physical non-BPS vertex operator, consider the half-BPS vertex operator
$|0\rangle_J$ corresponding to the long 
gauge-invariant super-Yang-Mills operator 
\eqn\verbps{ Tr (Z^J) }
with large $R$-charge $J$ where
$Z$ is the scalar at $x^m=0$ with $R$-charge $+1$ with respect to
a $U(1)$ direction of $SO(6)$. To be explicit, choose $Z=\phi_{12}$
where $\phi_{jk}$ are the six Yang-Mills scalars and
$J$ is the charge with respect
to the $U(1)$ generator $\half (R_1^1 + R_2^2 - R_3^3 - R_4^4)$.
The operator of \verbps\ is invariant under all $PSU(2,2|4)$ transformations
of \conjt\
except for the four translations $P_m$, the four $R$-symmetry generators
$(R_3^1, R_3^2, R_4^1, R_4^2)$, and the eight supersymmetry generators
$(q_3^\mu, q_4^\mu, \bar q^1_\md, \bar q^2_\md)$. Under these eight
bosonic and eight fermionic transformations, the operator of \verbps\
transforms in the same manner as in a Ramond-Ramond plane-wave background
when acted on with the eight bosonic and eight fermionic light-cone
oscillators \bmn.

To be more explicit, suppose that $(\Sigma_n)_j^k$ transforms $g(\s)$
as $\d g(\s) = e^{in\s} R_j^k g(\s)$. Then $(\Sigma_n)_3^1 |0\rangle_J$
is the vertex operator corresponding to the long gauge-invariant
Yang-Mills operator
\eqn\vernbps{ \sum_{m=1}^J Tr (Z^m ~\phi_{32} Z^{J-m}) 
e^{2\pi i n {m\over J}}.}
As in a plane-wave background,
this operator vanishes by cyclicity of the trace so one needs at least
two $\s$-dependent transformations to construct a physical states which
satisfies $L_0 -\bar L_0 =0$.
For example, 
$(\Sigma_{-n})_4^1 (\Sigma_n)_3^1 |0\rangle_J$
is the non-BPS vertex operator corresponding to the long
gauge-invariant Yang-Mills
operator
\eqn\vernbpst{ \sum_{m=1}^J Tr (\phi_{42} Z^m ~\phi_{32} Z^{J-m}) 
e^{2\pi i n {m\over J}}.}

The spectrum of these non-BPS operators is easily computed using the
$PSU(2,2|4)$ algebra. For example, $[D- J, R_3^1]=R_3^1$ and
$[D- J, R_4^1]=R_4^1$ where $D$ is the dilatation generator.
So the state 
$(\Sigma_{-n})_4^1 (\Sigma_n)_3^1 |0\rangle_J$ has eigenvalue $D-J=2$
which is independent of $n$. This agrees with the expected result at
zero `t Hooft coupling since the large $R$-charge formula
for the eigenvalue of the $n^{th}$ oscillator mode is
\eqn\formula{(D-J)_n = \sqrt{1 + {{4\pi g_s N}\over J} n^2}}
which is independent of $n$ when $g_s N=0$.

\subsec{Scattering amplitudes and open-closed duality}

If the topological action $S_{top}$ of \topa\ describes the zero-radius
limit of the \ads\ superstring, the \ads\ superstring at infinitesimal
radius $r$ should be described by the action 
\eqn\infi{S_r = S_{top} + r^2 S_{AdS}}
where $S_{AdS}$ is the vertex operator for the
radius modulus and is also the original \ads\ action of \class. Since
$S_{top}$ and $S_{AdS}$ are both invariant under the BRST transformation
generated by \brstp, \infi\ is also BRST invariant.\foot{Using the
previous proposal of $S_{top}$ based on the fermionic coset, such a
perturbation of $S_{top}$ would not be allowed since the topological
and \ads\ BRST operators were different.} Note that one could also
consider the action $S_r = t S_{top} + r^2 S_{AdS}$ where $t$
is a constant, but since
$S_{top}$ is BRST-trivial, the theory must be independent of the value
of $t$.

The Maldacena conjecture predicts that perturbative superstring
scattering amplitudes computed in the background of \infi\ should
coincide with perturbative correlation functions of gauge-invariant
super-Yang-Mills operators at small 't Hooft coupling.
Although it is not yet known how to compute topological string
amplitudes in the background of \infi, a handwaving argument
will be sketched based on open-closed topological duality that
such amplitudes should agree with the analogous super-Yang-Mills
computations. If this handwaving argument could be made rigorous,
it would provide a proof of the Maldacena conjecture at small
't Hooft coupling.

The handwaving argument is closely related to ideas in \gaiotto\ and
\gopa\oog\ 
which describe open-closed topological duality in the context of the
Kontsevitch model and Chern-Simons theory. The action $S_{top}$
of \topa\ describes a closed topological string theory, and one can 
define an open topological string theory by placing $M$ $D3$
branes at the boundary of $AdS_5$. As usual, the $D_3$ brane
boundary conditions are Dirichlet for the $(x^4, ..., x^9)$ variables,
Neumann for the $(x^0, ..., x^3)$ variables, and 
\eqn\purebdy{\lh^\ah = (\g_{0123})_\a^\ah \l^\a, \quad
\ww_\ah = (\g_{0123})_\ah^\a w_\a, }
for the pure spinor variables. Furthermore, the fermionic
boundary conditions imply that $J^\ah = (\g_{0123})^\ah_\a \bar J^\a$,
so the BRST operator satisfies $Q_L=Q_R$ on the boundary.

As discussed at the end of subsection (3.3), \purebdy\ implies
that $(\eta\l\lh) = \l\g^4\l=0$, so one needs to introduce
non-minimal variables on the boundary. These non-minimal variables
turn the zero mode measure factor into the same measure factor as
in a flat background which is the $d=4$ dimensional reduction of
\eqn\zerofl{\langle (\l\g^m\t)(\l\g^n\t)(\l\g^p\t)(\t\g_{mnp}\t)\rangle=1.}
One might be worried that the term
\eqn\singt{{{(\l\g_a)_\a\eta^{\a\ah}(\lh\g_b)_\ah}\over{2(\eta\l\lh)}} 
J^a \bar J^b}
in the action of \topa\ becomes singular on the boundary where 
$(\eta\l\lh)=0$, but the numerator 
$(\l\g_a)_\a\eta^{\a\ah}(\lh\g_b)_\ah$ also vanishes on the boundary
where it is proportional to $\l\g^a \g^4 \g^b \l=0$.

The first step in the open-closed duality argument is that the only
physical open string states on the $M$ $D_3$ branes are massless
$U(M)$ ${\cal N}=4$ super-Yang-Mills states. It is clear that these
super-Yang-Mills states are in the spectrum since the vertex operator
$V=\l^\a A_\a (x,\t)$ is in the open string BRST cohomology when
$A_\a(x,\t)$ satisfies the $d=4$ dimensional reduction of the
$d=10$ linearized super-Yang-Mills equations of motion. However,
the absence of other states in the open string BRST cohomology remains
to be proven. Nevertheless, it is reasonable that there are no other physical
open string states since the $D_3$ branes on the $AdS_5$ boundary preserve
$PSU(2,2|4)$ invariance, so any other such states would have to preserve
${\cal N}=4$ $d=4$ superconformal invariance and transform in the 
adjoint representation of $U(M)$.

The next step in the argument is that the open string field theory
action given by
\eqn\osft{{\cal S}  = {1\over{g^2}} \langle V Q V + {2\over 3}V  V  V\rangle}
reproduces the ${\cal N}=4$ $d=4$ super-Yang-Mills field theory
action where $V$ is the off-shell open string field, $g$ is the square-root
of the closed string coupling constant $g_s$, and the zero-mode measure
factor in \osft\ is the $d=4$ dimensional reduction of \zerofl.
This step is reasonable since, as in the Chern-Simons topological string
\ref\witch{E. Witten, {\it Chern-Simons gauge theory as a string
theory}, Prog. Math. 133 (1995) 637, hep-th/9207094.},
one expects the Feynman diagrams of the open topological string to
reduce to the Feynman diagrams of the massless field theory. And as
shown in 
\superpm\schwarz, 
the $d=10$ super-Yang-Mills field theory action (or
its dimensional reduction) can be expressed as 
${\cal S}  = {1\over{g^2}} \langle V Q V + {2\over 3} V  V  V\rangle$ where
$V=\l^\a A_\a(x,\t)$, $A_\a(x,\t)$ is an off-shell $d=10$
spinor superfield,
$Q= \l^\a D_\a$, $D_\a$ is the $d=10$ supersymmetric derivative, and 
$\langle ~~\rangle$ is the zero mode measure factor of \zerofl.
Furthermore, it will be assumed that as in the Chern-Simons
topological string \witch, closed string states decouple from 
open string states and do not contribute to open topological string scattering
amplitudes.

So when $r=0$ in \infi, it has been argued that the open 
string field theory for $M$ $D_3$ branes at the boundary describes
$U(M)$ super-Yang-Mills theory with coupling constant $g=\sqrt { g_s}$.
The final step in the argument is that adding the $r^2 S_{AdS}$
perturbation to $S_{top}$ in \infi\ affects the open string field theory 
by shifting the 't Hooft coupling constant. This step has an analog in
the open-closed duality of \gaiotto\ where parameters of the closed string
background of topological gravity were shown to affect the open
string field theory by shifting parameters in the Kontsevitch matrix
model. 

The justification for this step is that insertion
of a closed string vertex operator at a puncture in an open
topological string amplitude can be replaced
by expanding the puncture into a hole and inserting an appropriate
D-brane boundary state \ref\gaiottoone{D. Gaiotto, N. Itzhaki and
L. Rastelli, {\it Closed strings as imaginary D-branes},
Nucl. Phys. B688 (2004) 70, hep-th/0304192.}\gaiotto. 
For an arbitrary closed string vertex
operator, the corresponding D-brane boundary state may be difficult
to construct. But for the closed string vertex operator $S_{AdS}$
which is $PSU(2,2|4)$ invariant, it seems reasonable to assume
that the corresponding $D$-brane boundary state is proportional
to a $D_3$ brane at the $AdS_5$ boundary. Note that the proportionality
constant $f(r)$ must go to zero when $r\to 0$ in order
to be consistent with the assumed decoupling of closed string states
from open string states in the topological string.
So inserting the closed string vertex operator $S_{AdS}$ at a puncture
in an open topological string amplitude should be equivalent to expanding
the puncture to a $D_3$ brane hole and multiplying by a factor of $f(r)$.

Perturbing the background from $S_{top}\to S_{top} + r^2 S_{AdS}$
is equivalent to inserting an exponential set of closed string vertex
operators, and for each open string diagram with $h$ holes and $p$
punctures, the scattering amplitude is proportional to 
\eqn\amppl{ (g^2 M)^h (r^2)^p}
where $(g^2 M)^h$ comes from the usual $(\l_{'t Hooft})^h$ factor in
the 't Hooft expansion. Replacing the punctures by $D$-brane holes and 
including the proportionality constant of $f(r)$, the open string scattering
amplitude with $H$ holes is proportional to 
\eqn\ampltw{\sum_{h+p = H} {{(h+p)!}\over{h!p!}}(g^2 M)^h (r^2 f(r))^p = 
(g^2 M + r^2 f(r))^H}
where the factor of 
${{(h+p)!}\over{h!p!}}$
comes from the different ways to split the $H$ holes into
$h$ holes and $p$ punctures.

So in the background of \infi, it has been argued that the open string
field theory for $M$ $D_3$ branes on the $AdS_5$ boundary describes
super-Yang-Mills theory where the `t Hooft coupling is shifted from
$g^2 M$ to $g^2 M + r^2 f(r)$. Note that if one could
show that $f(r)$ were equal to $r^2$,
this argument
would imply that the `t Hooft coupling is equal to $r^4$ when $M=0$.
So the relation $\l_{'t Hooft} =r^4$ would be valid both at small and large
radius.

\newsec{Conclusions and Discussion}

In the first half of this paper, it was shown that $(\eta\l\lh)$ is
in the BRST cohomology in an \ads\ background, 
which implies that the left and right-moving
pure spinor ghosts can be treated as complex conjugate variables.
This eliminates the need for non-minimal variables and simplifies
the zero-mode measure factor and $b$ ghost.

In the second half of this paper, a BRST-trivial version of the \ads\
action was constructed by gauge-fixing a ${ G}/{ G}$
principal chiral model where ${ G}=PSU(2,2|4)$. This topological
action was argued to describe the zero radius limit which is dual
to free super-Yang-Mills, and perturbing the topological action
by the vertex operator for the radius modulus was conjectured
to describe super-Yang-Mills at small 't Hooft coupling.

One possible method for proving this conjecture uses open-closed
topological string duality along the lines proposed in the previous
subsection. However,
a more direct method would be to compute the topological closed
string amplitudes and compare with the perturbative
Feynman diagrams of the
super-Yang-Mills field theory.
In \topthree, 
a connection was found between networks of Wilson lines constructed
from worldsheet gauge fields in the ${ G}/{ G}$ model and
the propagators and vertices of ${\cal N}=4$ super-Yang-Mills Feynman
diagrams. It would be very exciting if amplitude computations in
the topological model could be related to counting these Wilson line
networks in the ${ G}/{ G}$ model.

Although it is well-understood how to compute scattering
amplitudes with conventional
topological string theories, the topological model of \topa\
has some new features
which have not yet been studied. Unlike the usual topological strings
where the complex structure of the target spacetime is fixed, the
complex structure of the target spacetime in \topa\ is determined
dynamically by the pure spinors $(\l^\a,\lh^\ah)$
which choose a $U(5)$ subgroup of the (Wick-rotated) $SO(10)$
Lorentz group. This can be seen from the kinetic term for the $x$'s
in the topological action which, to quadratic order, is
$\int d^2 z (2 \eta\l\lh)^{-1}\eta^{\a\ah}(\l\g_a)_\a(\lh\g_b)_\ah
\p x^a \pb x^b.$
So classical instanton solutions satisfy
\eqn\instc{(\l\g_a)_\a \p x^a =0,\quad (\lh\g_a)_\ah \pb x^a=0,}
where $(\l\g_a)_\a$ determines which five complex components of
$\p x^a$ must vanish. 

Another new feature of the topological sigma model of \topa\ is that the 
ghost-number anomaly does not fix the number of unintegrated versus
integrated vertex operators. Since vertex operators can be multiplied
by inverse powers of $(\eta\l\lh)$ without spoiling BRST invariance,
one can construct unintegrated vertex operators of ghost-number zero
such as $V=(\eta\l\lh)^{-1} \l^\a \lh^\ah A_{\a\ah}(x,\t,\th)$.
It is unclear if the topological amplitude prescription should involve
both unintegrated and integrated vertex operators, or only unintegrated
vertex operators. Similarly, it is unclear if the genus $g$ topological
amplitude prescription requires integration over the moduli of genus $g$
Riemann surfaces.

In addition to describing the zero radius \ads\ limit, the topological
model of \topa\ can also be interpreted as a tensionless string in which all
massless and massive background fields are treated on equal footing.
Changing the 
target-space metric in the topological action is a 
BRST-trivial operation so,
as proposed by Witten, the topological 
model describes string theory in an ``unbroken phase'' in
which general covariance does not require an explicit metric \wittop\wittopq.

By giving background values to physical moduli,
one can perturb the topological model into non-topological
string theories which describe backgrounds that are asymptotically \ads\ but
are not necessarily $PSU(2,2|4)$ invariant.
For example, perturbing with the vertex operator for the radius modulus
deforms the topological action into the $PSU(2,2|4)$-invariant
\ads\ action of \class, but perturbing with other physical moduli will lead
to superstring backgrounds which are asymptotically \ads\ but which
are not $PSU(2,2|4)$ invariant. 

In some sense, these asymptotically \ads\ backgrounds are more natural
backgrounds for the pure spinor formalism than asymptotically flat
backgrounds. In asymptotically \ads\ backgrounds, the worldsheet
action can always be constructed from the Metsaev-Tseytlin
coset $g\in {{PSU(2,2|4)}\over
{SO(4,1)\times SO(5)}}$ even though the action is not necessarily
invariant under the global $PSU(2,2|4)$ isometries $\d g = \Sigma g$.
Furthermore, the BRST operator in these backgrounds
always acts geometrically as
$Q g = g(\l^\a T_\a + \lh^\ah T_\ah)$ and there is no need
to introduce non-minimal variables. And in the limit where the
radius goes to zero, the topological \ads\ pure spinor action 
and BRST operator can
be derived by gauge-fixing a ${ G}/{ G}$ principal chiral
model. This contrasts with the pure spinor formalism in a flat
background which has not yet been derived in a simple manner from
gauge fixing.

\vskip 1cm
{\bf Acknowledgments:}
I would like to thank Y. Aisaka, R. Gopakumar, P. Howe, 
J. Maldacena, A. Mikhailov, N. Nekrasov, A. Polyakov, 
W. Siegel, C. Vafa, B.C. Vallilo, H. Verlinde
and E. Witten for useful discussions, and
CNPq grant 300256/94-9 and FAPESP grant 04/11426-0
for partial financial support.

\listrefs

\end